\documentclass[superscriptaddress,amsmath,amssymb,aps,onecolumn,pra]{revtex4-2}

\def\eps{\varepsilon}

\usepackage{amssymb}
\usepackage{amsmath}
\usepackage{color}
\usepackage{soul}
\usepackage[utf8]{inputenc} 
\usepackage[T1]{fontenc}    
\usepackage{graphicx}       
\usepackage[margin=2.5cm]{geometry}
\usepackage{float}

\begin{document}

\title{Opinion formation in Wikipedia Ising networks $^{\dagger}$}


\author{Leonardo~Ermann}
\affiliation{Departamento de F\'{\i}sica Te\'orica, GIyA,
         Comisi\'on Nacional de Energ\'{\i}a At\'omica.
           Av.~del Libertador 8250, 1429 Buenos Aires, Argentina}
           
\author{Klaus~M.~Frahm}
\affiliation{Univ Toulouse, CNRS, Laboratoire de Physique Th\'eorique,  
Toulouse, France}
\author{Dima~L.~Shepelyansky}
\email{dima@irsamc.ups-tlse.fr}
\affiliation{Univ Toulouse, CNRS, Laboratoire de Physique Th\'eorique,  
Toulouse, France}


\begin{abstract}
 We study properties of opinion formation
  on Wikipedia Ising Networks. Each Wikipedia article
  is represented as a node and links are formed by citations of
  one article to another generating a directed network
  of a given language edition with millions of nodes.
  Ising spins are placed at each node
  and their orientation up or down is determined by a majority vote
  of connected neighbors. At the initial stage there are only 
  a few nodes from two groups with fixed competing opinions up and down
  while other nodes are assumed to have no initial opinion with no
  effect on the vote. The competition of two opinions is modeled by
  an asynchronous Monte Carlo process converging to a spin polarized 
  steady-state phase. 
  This phase remains stable with respect to small fluctuations
  induced by an effective temperature of the Monte Carlo process.
  The opinion polarization at the steady-state provides
  opinion (spin) preferences for each node. In the framework of 
  this Ising Network
  Opinion Formation model we analyze the influence and competition between
  political leaders, world countries and social concepts.
  This approach is also generalized to the competition between 
  three groups of 
  different opinions described by three colors, for example 
  Donald Trump, Vladimir Putin, Xi Jinping or USA, Russia, China
  within  English, Russian and Chinese editions of Wikipedia of March 2025.
  We argue that this approach provides a generic description of
  opinion formation in various complex networks.
\end{abstract}
  



\maketitle

\section{Introduction}

The process of opinion formation in human society gains higher and 
higher importance with the development of social networks
which start to produce an important impact on political views
and elections (see e.g. \cite{soc1,soc2}).
The statistical properties of such social networks
typically have a scale-free structure
as reviewed in   \cite{fortunato09,dorogovtsev10}.
Various voter models have been proposed and studied by different groups
with a development of physical concepts and their applications
to sociophysics
\cite{fortunato09,galam82,galam86,sznajd00,sood05,watts07,galam08,schmittmann10}.

Recently, we proposed the Ising Network Opinion Formation (INOF) model
and analyzed its applications
to Wikipedia networks for 6 language editions of 2017 \cite{inof24}. 
This model allows to determine opinion polarization for all 
Wikipedia articles (or nodes) induced by two groups of nodes
with fixed opposite opinions (red or blue, spin up or down). 
In this INOF model the initial two groups of one or a few nodes 
have fixed opposite opinions represented by spin up (red color) or
spin down (blue color). All other nodes have initially an undecided opinion
(spin zero or white color). The formation of a steady-state opinion of each
node emerges as a result of an asynchronous Monte Carlo process
in which an opinion of a given node $i$ 
is determined by a majority vote of his friends 
presented by spins up or down or zero from
all network nodes $j$ that have links to node $i$. 
Such spin flips, induced by local majority votes,
are done for all nodes without repetitions in random order over all $N$ nodes. 
These procedure is repeated up to convergence to a steady-state for 
a sufficiently long time $\tau$ 
and corresponds to a particular random pathway realization 
for the order of spins to be flipped. 
Finally, an average over a high number of pathways realizations is done 
to obtain averages and distributions of the opinions for nodes 
or the whole network (see next section for more technical details). 
A somewhat similar procedure is used in the studies
of problems of associative memory (see e.g. \cite{memory1,memory2})
even if there are significant differences from the INOF model
due to the absence of certain fixed nodes and other initially white nodes 
and the use 
of positive/negative transition elements  between nodes while all of them 
are positive for the INOF case considered here. 

The mathematical and statistical properties of certain models
of Ising spins om complex networks have been analyzed in
\cite{dorogovtsev,bianconi}. However, these works
studied very different aspects of such networks compared to
those discussed in our work.

We also note that the INOF approach is generic and
can be applied to various directed networks.
In particular, it has also been applied to the analysis of fibrosis progression
in the MetaCore network of protein-protein interactions \cite{fibrosis2}.
A similar approach, without white nodes, 
was used to study the competition of dollar and possible BRICS currencies
in the world trade network \cite{brics2024}.

In this work, we extend the studies of the INOF model \cite{inof24} 
to more recent Wikipedia networks collected either 
in October 2024 or collected at 20 March 2025 
and to new specific initial groups with fixed opinions. 
For example, we analyze the 
competition between 
Apple Inc. vs. Microsoft, Donald Trump vs. Vladimir Putin  and others. 
Furthermore, the competition of three entries is also considered for 
several cases. Thus we analyze the competition between three groups of
Donald Trump, Vladimir Putin, Xi Jinping,
and compare the results with the case of competition
between 3 related countries  USA, Russia, China.
We also show that the INOF approach allows to
analyze the interaction and influence of such social
concepts as Liberalism, Communism, Nationalism.
In this work we consider two types of vote contributions
(determined by two representations of $V_{ij}$ in (\ref{eqz}), see below)
while only one of them was considered in \cite{inof24}.
And finally, we also examine the effects of fluctuations
appearing as a result of a certain effective temperature in the 
voting process. 

Wikipedia networks have rather exceptional
features as compared to other networks:
the meaning of their nodes is very clear, 
they represent all aspects of nature and human activity
and the presence of multiple language editions
allows to analyze various cultural views of humanity.
A variety of academic research of Wikipedia with analysis 
of different aspects of nature and society
was reviewed in \cite{wikiacad1,wikiacad2,wikiacad3,wikiacad4,wiki25}.
Therefore we hope that the INOF approach to
Wikipedia Ising networks (WIN) will find multiple and divers 
applications.

The article is composed as follows:
Section 2 describes the INOF model and the used Wikipedia data sets,
 Section 3 presents results for the confrontation of
 opinions for two groups of entries,
 Section 4 presents results for a three groups contest,
 Section 5 analyzes effects of fluctuations
 induced by an effective temperature and Section 6
 contains discussion and conclusion. Finally, the Appendix provides 
 some additional Figures and data. 

 We note that the preliminary results of this research
have been presented at the contributed talk of authors
at Wiki Workshop 2025 on May 22
(the abstract of 3 pages is available at \cite{wiki25}).

\section{Model description and data sets}

In this work, we use mostly  three very recent Wikipedia editions 
(English EN, Russian RU, Chinese ZH) collected at 20 March 2025
with the number of network nodes/articles $N=6969712,\,2035086,\,1468935$ 
and the number of links 
$N_{\ell} = 190031938,\,44188839,\,21160179$ for 
EN, RU, ZH respectively. For certain cases, we also use 
the English (EN) and French (FR) Wikipedia network 
collected at 1 October 2024 and already used previously (see \cite{wiki25} and Refs. therein) 
with $N = 6891535, 2638634 $ and $N_{\ell} = 185658675, 76118849 $ 
for EN (2024), FR (2024) respectively.

Wikipedia articles correspond to network nodes and citation from a given 
article $j$ to another article $i$ correspond to a directed link 
with the adjacency matrix element $A_{ij}=1$ 
(and $A_{ij}=0$ in absence of a link from $j$ to $i$);
multiple citations from $j$ to $i$ are considered as only one link. 
The above definition of link directions of $A_{ij}$ 
corresponds to those used in our previous works, see e.g. \cite{wiki25} and Refs. therein,
and we keep it here.
Then the matrix of Markov transitions is defined by 
$S_{ij}=A_{ij}/k_{j}$ where $k_{j}=\sum_i A_{ij}$ is a number of 
out-going links
from node $j$ to any other node $i$ (such that $A_{ij}\neq 0$); 
for the case of dangling nodes without out-going links 
(i.e. with $k_j=0$), we simply define $S_{ij}=1/N$ 
implying the usual column sum normalization $\sum_i S_{ij} =1$ for all $j$.
For later use, we also introduce the modified matrix $\tilde S_{ij}$ 
which is identical to $S_{ij}$ for $k_j>0$ and with $\tilde S_{ij}=0$ 
for dangling nodes. 

Usually, in other typical types of network studies (see e.g. \cite{wiki25}) 
one introduces the Google matrix of the network  defined as
 $G_{ij}= \alpha S_{ij} +(1-\alpha)/N$ 
where alpha is the damping factor with the standard value 
$\alpha =0.85$ \cite{brin,meyer}.
Here the network nodes can be characterized by 
the PageRank vector which is the eigenvector of
the Google matrix $G$ \cite{brin,meyer}
with the highest eigenvalue $\lambda=1$, i.e. 
$G P = \lambda P = P$, and the damping factor $\alpha<1$ ensures 
that this vector is unique and can be computed efficiently. 
Its components $P(i)$ are positive and normalized to unity 
(${\sum_{i=1}}^N P(i) =1$). The network nodes $i$ can be ordered 
by monotonically decreasing probabilities $P(i)$ which 
provides the PageRank index $K$ with 
with highest probability at $K=1$ and smallest at $K=N$. Some 
results for the PageRank vector and its index for recent Wikipedia editions 
of 2024 can be found at \cite{wiki25} and Refs. therein. However, in this work, we do 
not use the Google matrix neither the PageRank and focus mostly 
on the modified matrix $\tilde S$ and also the adjacency matrix $A_{ij}$  to define 
an asynchronous Monte Carlo process. 

As in \cite{inof24} a few selected nodes (wiki-articles) have 
assigned fixed spin values
$\sigma_l=-1$ blue e.g.
for {\it Microsoft}
and $\sigma_k=1$ red for {\it Apple Inc.}.
These specific spin nodes always keep their polarization.
All other nodes $i$ are initially assigned with a white color
(or spin $\sigma_i=0$) and have no definite initial opinion. 
However, once they acquire a different 
color red or blue (spin value $\sigma_i=\pm 1$) during the 
asynchronous Monte Carlo process they can flip only between $\pm1$ and
cannot change back to the white opinion. 

To define the asynchronous Monte Carlo process, 
we choose a random spin $i$ among the non-fixed set of spins,
and compute its influence score from ingoing links $j$:
\begin{equation}
	Z_i = \sum_{j \neq i} \sigma_{j} V_{ij} ,
	\label{eqz}
\end{equation}
where the sum is over all nodes $j$ linking to node $i$.
Here $V_{ij}$ is the element of the vote matrix, 
defined by one of two options:
$V_{ij}=A_{ij}$ (the adjacency matrix element, option OPA), 
or $V_{ij}={\tilde{S}}_{ij}$ (the modified Markov transition 
matrix element, option OPS). 
For the OPS option, the matrix  $\tilde S_{ij}$ is used, 
in which columns corresponding to dangling nodes contain 
only zero elements, ensuring these nodes do not contribute to $Z_i$.
We discuss both options OPA and OPS with a primary focus on the OPS case.

In Eq.~(\ref{eqz}) $\sigma_j=1$ if the spin of node $j$ is oriented up 
(red color), or $\sigma_j=-1$ if it is oriented down (blue color), or
$\sigma_j=0$ if the node $j$ has no opinion 
(if it has still its initial white value). 
After the computation of $Z_i$, 
the spin $\sigma_i$ of node $i$,  is updated:
it becomes $\sigma_i=1$ if $Z_i>0$, $\sigma_i=-1$ if $Z_i<0$, 
and remains unchanged if $Z_i=0$. 
This operation is repeated for all non-fixed nodes $i'$ 
following a predetermined random order (shuffle)
such that there is no repetition at this level and 
each spin is updated only once. Note that due to the possibility of $Z_i=0$ it 
is possible that a node $i$ keeps his initial white value $\sigma_i=0$. 
After the update of $\sigma_i$, the modified value of $\sigma_i$ 
is used for the computation of $Z_{i'}$ of subsequent values 
$\sigma_{i'}$. 

One full pass of updating all non-fixed spins constitutes 
a single time step, $\tau=1$. 
The procedure is then repeated for subsequent time 
steps $\tau=2,3,\ldots$ using a new random shuffle for 
the update order at each step.
We find that the final steady-state is reached after 
$\tau \approx 20$ steps with only a very small number of spin flips 
in the $\tau=20$. 
There is a certain fraction of nodes
that remain white for $\tau \geq 20$ 
which we attribute to their presence in isolated communities 
(about 12\% for EN 2025, 15\% for RU 2025 and 30\% for ZH 2025 
and 10\% for FR 2024). 
These nodes are not taken into account when determining 
the opinion polarization of other nodes and all statistical quantities 
such as averages, fractions and histograms are computed with 
respect to the set of non-white nodes. 
We point out that compared to the usual case of Wikipedia networks
the size of the configuration space of the INOF model is 
drastically increased to $2^N$ instead of $N$. 

The physical interpretation of the OPA case corresponds to the 
situation where a node $j$ gives an unlimited number of votes to 
the nodes $i$ to which he has links 
while for the OPS case the node $j$ has only a limited vote capacity 
(since  the total probability in column $j$ is normalized to unity).
Therefore these two options OPA and OPS describe two different
possibilities for the voting process.
We note that due to a misprint in \cite{inof24}
the analysis was performed for OPA case and not with OPS one
as it is declared in  \cite{inof24}.

Repeating this asynchronous Monte Carlo process, with 
the same initial condition and different 
random orders (or pathways) for the spin flip defined by the rule (\ref{eqz}), 
we obtain various random realizations
leading to different final steady-state distributions in each case. 
Using this data we perform an average over up to $N_r = 10^5$
pathway realizations ($N_r=10^6$ for the case of FR 2024 to obtain 
a reduced statistical error for this case; see below) 
that provides 
an average opinion polarization $\mu_i$ of a given spin (node, article). 
The further average of $\mu_i$ over all (non-white) network nodes gives 
the global polarization $\mu_0$ 
with a deviation $\Delta \mu_i =\mu_i - \mu_0$ for each article.
This deviation $\Delta \mu_i$ represents the opinion preference of
a given article $i$ to red or blue entries as compared to
the average global Wikipedia opinion $\mu_0$.
The set of white nodes in the final steady-state distribution
contains about 10\% - 30\% of the total number of nodes 
(30\% only for ZH Wiki2025 and at most 15\% for the 
other cases) and this set is extremely stable 
with respect to different pathway realizations and also with 
respect to the different choices of initial fixed nodes. 
These white nodes are not taken into account in the computation of $\mu_0$ 
and $\mu_i$ is only computed for non-white nodes (those which have nearly 
always either red or blue values depending on the pathway realization). 

The voting process for the case of a competition between 
three groups of entries is an extension 
of this procedure and its details are  explained later.

\section{Results for competition of two groups of entries}

\subsection{Comparison of OPA and OPS}

The relaxation of Ising spins to the steady-state in
Wiki2024, Wiki2025 networks is very similar to those found for Wiki2017
networks studied in \cite{inof24} (see e.g. Figure 1 there).
Thus the steady-state is reached at $\tau \geq 20$,
the fractions of red and blue nodes at the final state
are concentrated mainly at all red or at all blue nodes. 

\begin{figure}[H]
\begin{center}
\includegraphics[width=0.7\textwidth]{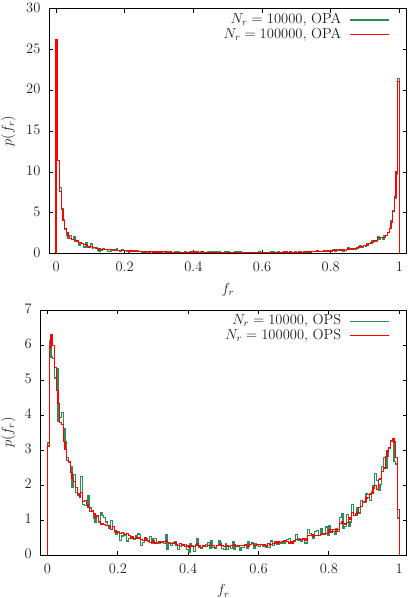}  
\caption{\label{fig1}
Probability density $p(f_r)$ of fraction of red nodes $f_r$ for SCCI for EN Wiki2025 versus $f_r$ 
using the matrix $A$ (OPA case, top panel) or the matrix 
$\tilde S$ (OPS case, bottom panel). The value of $f_r$ 
for a given pathway realization gives the fraction of red 
outcome of network nodes
for Socialism/Communism 
(fraction computed with respect to all non-white network nodes). 
The histograms show the distribution of $f_r$ with respect to
all $N_r$ different random pathway realizations and all nodes.
The normalization is fixed by $\int_{0}^1 p(f_r)\,df_r=1$ with 
bin width $0.005$ (i.e. 200 bins in the full interval for $f_r\in[0,1]$).
The red (green) curves correspond to $N_r=100000$ ($N_r=10000$). 
Note that the distribution $\tilde p(f_b)$ for blue outcome for
Capitalism/Imperialism (not shown in the figure) 
can be obtained by symmetry $\tilde p(f_b)=p(1-f_b)$ since $f_r+f_b=1$. 
}
\end{center}
\end{figure}

For each pathway realization we compute the fraction of red nodes 
$f_r$ as the ratio 
$f_r=n_r/(n_r+n_b)$ where $n_r$ ($n_b$) is the number of red (blue) 
nodes in the network for this pathway realization at $\tau=20$. In particular, 
white nodes are not taken into account in this fraction and 
the corresponding fraction of blue nodes is simply by symmetry 
$f_b=n_b/(n_r+n_b)=1-f_r$. 
We show examples for the probability density of $f_r$ (histogram 
for many different pathway realizations)
in Figure~\ref{fig1} for two groups of entries  {\it socialism, communism} (red) 
vs. {\it capitalism, imperialism} (blue)  SCCI
for the two cases OPA and OPS of EN Wiki2025.

We note that the probability distribution $p(f_r)$ in Figure~\ref{fig1} is
obtained as a histogram using $N_r$ values of $f_r$ for the 
obtained for the different pathway realizations. The shape of 
distributions for OPA and OPS cases are similar with strong 
maxima close to $f_r\approx 0$ and $f_r\approx 1$ but for the OPA case the 
distribution has sharper peaks. 
In both cases, it is very likely that either one or the other opinion is 
a strong winner for a given random pathway realization but this effect 
is somewhat stronger for the OPA case where the peak maxima are 
roughly four times larger than for the OPS case. 
Furthermore, the comparison of the two curves for $N_r=10^4$ and 
$N_r=10^5$ indicates the reduction of statistical fluctuations with 
increasing $N_r$. 

In this work, we compute for many cases and situations 
the average spin polarization $\mu_i$ of a node $i$ with respect 
to the $N_r$ random pathway realizations by the equation 
$\mu_i=(n_r(i)-n_b(i))/(n_r(i)+n_b(i))$ 
where $n_r(i)$ ($n_b(i)$) is the number of red outcome of node $i$ 
(blue outcome of node $i$) for the $N_r$ different random pathway 
realizations of the Monte Carlo procedure. 
Here, for the non-white nodes $i$ we typically 
have $n_r(i)+n_b(i)\approx N_r$ and the number of white outcome $n_w$ 
of these nodes is very small $n_w(i)=N_r-(n_r(i)+n_b(i))\ll N_r$ (it is not 
always exactly zero due a limited iteration time $\tau=20$ in the Monte 
Carlo procedure). 
For the fraction of white nodes (about 12\% for EN Wiki2025), 
we have typically $n_w(i)\approx N_r$ 
and we do not compute $\mu_i$ for these nodes for which $\mu_i$ is 
either not even defined (if $n_b(i)+n_r(i)=0$) or has a large statistical 
error (if $0<n_b(i)+n_r(i)\ll N_r$). 
Once $\mu_i$ is known for a given non-white node, we also compute 
the difference $\Delta\mu_i=\mu_i-\mu_0$ where $\mu_0$ is the global 
network average of $\mu_i$ over all (non-white) nodes $i$. This difference 
represents the opinion preference of the node $i$ in comparison to 
the average global network opinion $\mu_0$ and it will be used in several 
of the subsequent figures and tables to present data for the case of a 
two-way competition. 

\begin{figure}[H]
\begin{center}
\includegraphics[width=0.7\textwidth]{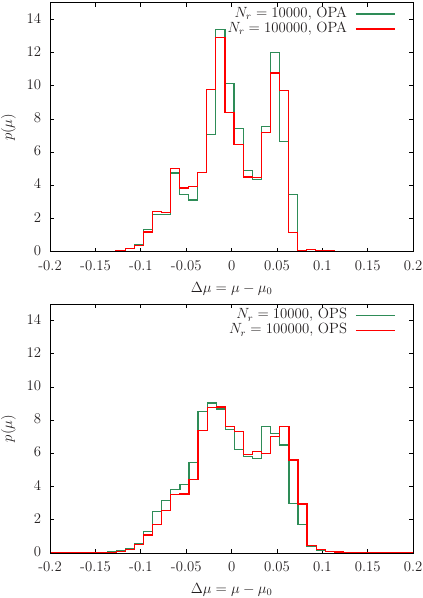}
  \caption{\label{fig2}
Probability density $p(\mu)$ of polarization $\mu$ for 
SCCI for EN Wiki2025 versus $\Delta\mu=\mu-\mu_0$ 
using the matrix $A$ (OPA case, top panel) or the matrix 
$\tilde S$ (OPS case, bottom panel). Here a red (blue) outcome with 
$\mu_i\approx 1$ ($\mu_i\approx -1$) corresponds to Socialism/Communism 
(Capitalism/Imperialism). 
The histograms show the distribution of $\mu_i$ for all 
non-white nodes $i$ in the network with 
with $n_r(i)+n_b(i)\approx N_r$ and white nodes with $n_r(i)+n_b(i)\ll N$ 
are not taken into account. 
The normalization is fixed by $\int_{-1}^1 p(\mu)\,d\mu=1$ with 
bin width $0.01$ (i.e. 200 bins in the full interval for $\mu_i\in[-1,1]$). 
Values of $p(\mu)$ outside the shown interval $[-0.2,0.2]$ are zero on 
graphical precision. The red (green) curve corresponds to 
$N_r=100000$ ($N_r=10000$). The value $\mu_0$ is the average of
$\mu_i$ with respect to all (non-white) nodes $i$ 
(computed for  $N_r=100000$) and has the values $\mu_0=-0.02333$ for 
OPA and $\mu_0=-0.10886$ for OPS.
}
\end{center}
\end{figure}

In particular the probability density $p(\mu)$ of $\mu_i$ for the SCCI 
case of EN Wiki2025 (red for Socialism/Communism and blue for 
Capitalism/Imperialism) 
is shown in Figure~\ref{fig2} for both cases OPA and OPS.

Here the $p(\mu)\,\delta\mu$ represents the fraction of (non-white) 
network nodes $i$ with $\mu\le\mu_i\le \mu+\delta\mu$ for some small 
value of $\delta\mu$ (bin width $\delta\mu=0.01$ in the histograms of 
Figure~\ref{fig2}). 
In global the profiles of OPA and OPS distributions are similar
even if there are certain differences related to different voting procedures.
As in Figure~\ref{fig1}, the curves of Figure~\ref{fig2} at $N_r=10^5$ seem 
to be rather stable with respect to statistical fluctuations as the comparison 
with the curves for $N_r=10^4$ shows. In Figure~\ref{fig2}, these 
fluctuations are a bit higher as in Figure~\ref{fig1} the computation of 
$\Delta\mu_i$ indeed requires at least $N_r=10^5$ while $N_r=10^4$ is 
not really sufficient. 

The statistical accuracy of the polarization  $\mu_i$ of a given node $i$
computed for $N_r$ realizations can be obtained in the following way:
we know that $\mu_i=\langle\sigma_i\rangle$ where $\sigma_i=\pm 1$ 
is the spin value after $\tau=20$ Monte Carlo iterations for one 
specific pathway realization and 
$\langle\sigma_i\rangle$ is simply the average of $\sigma_i$ with respect 
to the $N_r$ random pathway realization. Since $\sigma_i^2=1$, we find that 
$\langle \sigma_i^2\rangle=1$ and 
the variance of $\sigma_i$ is simply Var$(\sigma_i)=1-\mu_i^2$. 
From this, we obtain the statistical error of 
the average $\mu_i$ as $\sqrt{(1-\mu_i^2)/(N_r-1)}\sim 1/\sqrt{N_r}$
for $\mu_i\approx 0$ (case of ``largest'' error). For $N_r=10^5$ 
this gives a theoretical statistical error of $\mu_i$ being 
$\approx 0.003$ for $\mu_i\approx 0$ 
(and a slightly reduced error by a factor $1-\mu_i^2$ if $\mu_i\neq 0$). 

It is also possible to compute the statistical error quite accurately 
by a more direct numerical computation. For this we divide the full data of 
global $N_r=10^5$ pathway realizations 
into 100 samples with reduced $N_r=10^3$ and 
compute for each sample partial averages 
$\mu_i$, $\mu_0$ and $\Delta\mu_i$ over 
the reduced value of $N_r=10^3$. 
This provides in particular partial averages $\mu_0$ with quite 
significant fluctuations between the samples. Using the 100 partial 
average values $\Delta\mu_i$ for each sample 
it is straightforward to determine the statistical error of $\Delta\mu_i$ as 
$\approx \sqrt{\langle \Delta\mu_i^2\rangle-
\langle \Delta\mu_i\rangle)^2/(100-1)}$ where $\langle\ldots\rangle$ 
is the simple average over the 100 samples (of the partial sample averages). 
It turns out that typical error values obtained in this way 
(at global $N_r=10^5$ and for SCCI of EN Wiki2025 and a few other cases) 
are closer to $0.0008-0.0015$ which is about $2-3$ times smaller than the 
theoretical error $\approx 0.003$. This reduction is apparently due 
to rather strong 
correlations between $\mu_i$ and $\mu_0$ which are likely to have 
statistical fluctuations in the same direction (partial sample averages of 
both are likely to be ``large'' or ``small'' at the same time). 
These correlations are also visible in the two peak distribution of 
Figure~\ref{fig1} showing that in one given pathway realization it is rather 
likely to have 
either $f_r\approx 1$ (or $f_r\approx 0$) corresponding to a majority 
of network nodes $i$ with either $\sigma_i=1$ (or $\sigma_i=-1$ respectively). 
We note that the numerical statistical error of $\mu_0$, which is obtained 
as byproduct of the above procedure, is $\approx 0.002-0.003$ which is closer 
to the theoretical error. 

We also remind that to obtain $f_r$ in Figure~\ref{fig1}, we first perform 
an effective average over the network for a {\em fixed random pathway realization} 
(counting the number or fraction of red nodes in the network which is 
also the network average of $(\sigma_i+1)/2$) and then in 
Figure~\ref{fig1} we show histograms of this quantity using $N_r$ values 
of $f_r$ obtained from the $N_r$ random pathways. 
For the data shown in Figure~\ref{fig2} 
this is essentially the other way round: first we compute $\mu_i$ 
as the average of $\sigma_i$ over the $N_r$ random pathway realizations 
and then we compute a histogram using the obtained $\mu_i$ values for 
all (non-white) network nodes $i$. 

\begin{figure}[H]
\begin{center}
  \includegraphics[width=0.65\textwidth]{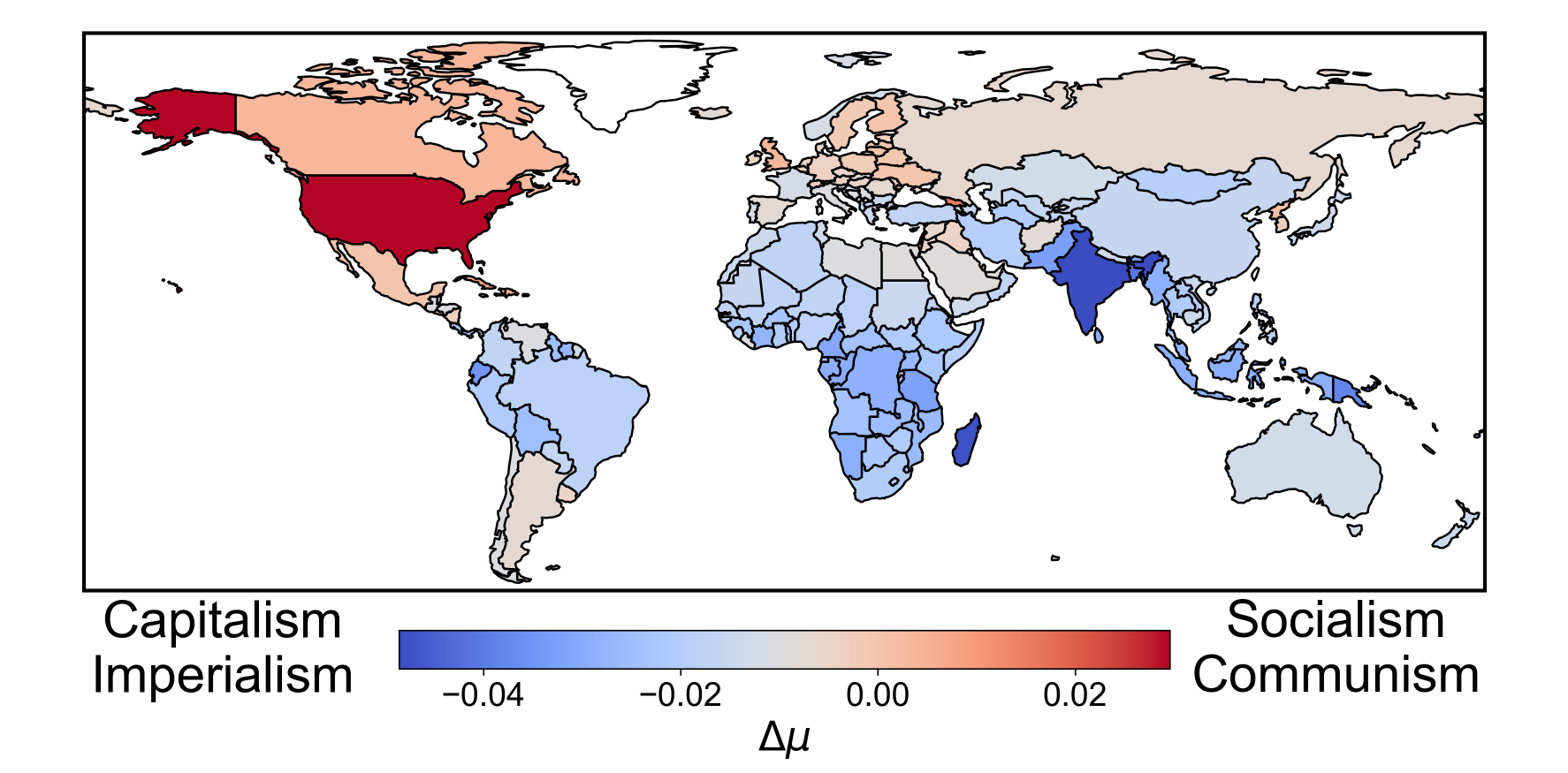}
  \includegraphics[width=0.65\textwidth]{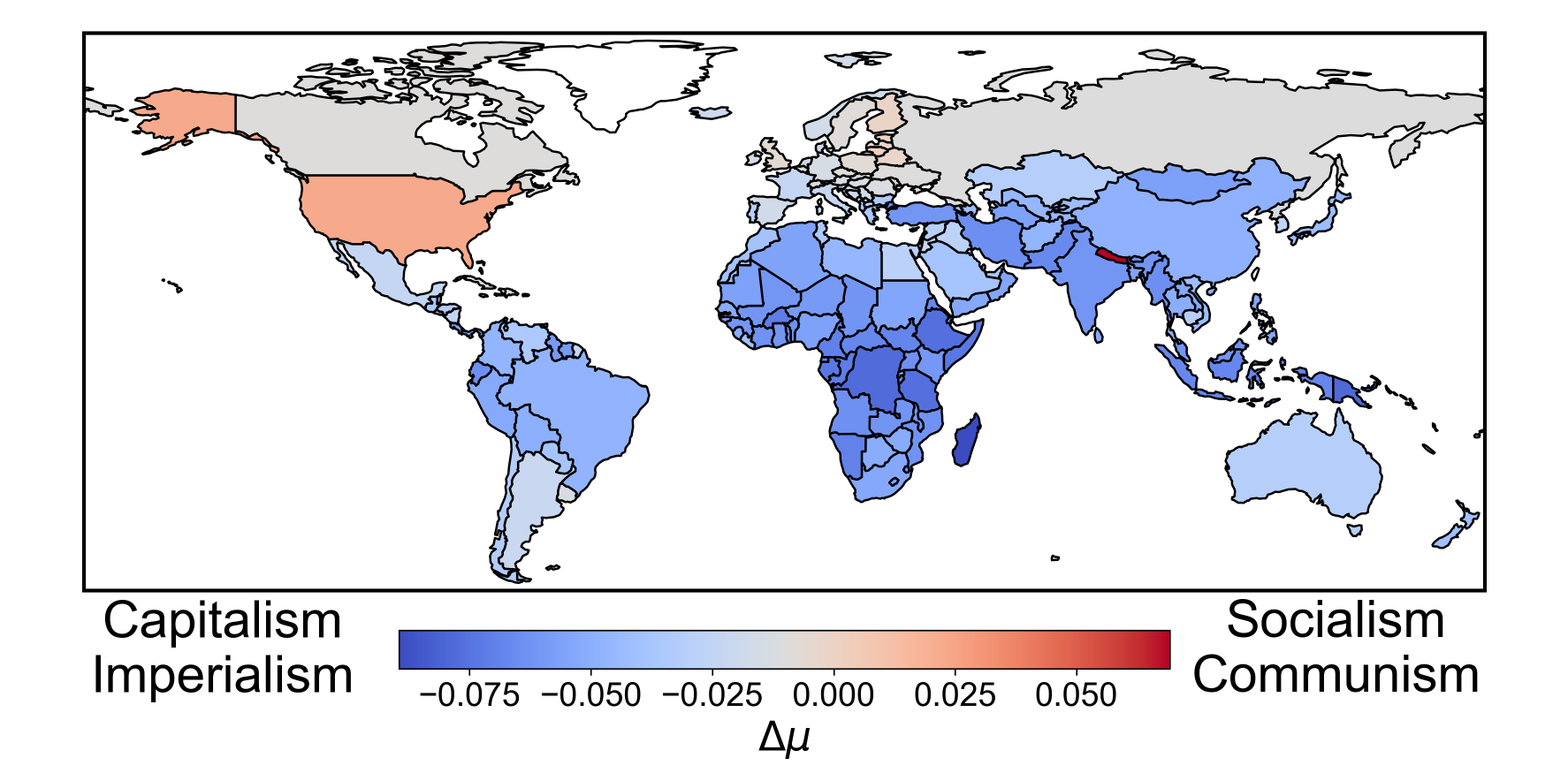}
\caption{\label{fig3}
Opinion polarization of world countries
for {\it socialism, communism} ($\Delta \mu >0$) vs.
{\it capitalism, imperialism} ($\Delta \mu <0$) for
OPA case (top panel with $\mu_0=-0.023$) and
OPS case (bottom panel with $\mu_0=-0.109$)  for EN Wiki2025
and 197 countries.
}
\end{center}
\end{figure} 

The world map of countries (taken as Wikipedia article of a given country)
with their polarization values  $\mu_i$ and $\Delta \mu_i = \mu_i - \mu_0$
is shown by color in Figure~\ref{fig3} for SCCI from
EN Wiki2025. For both cases OPA and OPS the global polarization $\mu_0 < 0$
is in favor of {\it capitalism, imperialism} 
but for OPS $|\mu_0|$ is by a factor 4 higher compared to the OPA case.
The global distribution of colors on the world map is
qualitatively similar for both OPA and OPS cases.
However,  in the OPS case the preference to {\it capitalism, imperialism} 
is significantly more pronounced with almost all countries of Africa,
dominance in Latin America, India, China and
other countries in a south of Asia. In Europe lowest 
$\Delta \mu$ values
are for France, Italy being however significantly larger 
compared to those of China, India, Africa. For Russia $\Delta \mu$ is 
closer to zero which is similar to the cases of Canada and Germany.
USA has clearly a positive value of $\Delta \mu$.
We note that similar opinion polarization of countries
have been seen for EN Wiki2017 in \cite{inof24}.

After the comparison of OPA and OPS cases we conclude that they
provide qualitatively similar results
even if there are quantitative differences between these two voting options.
In the following, we present results mainly for the OPS case.

\subsection{Competition Apple Inc. vs. Microsoft}

\begin{figure}[H]
\begin{center}
\includegraphics[width=0.65\textwidth]{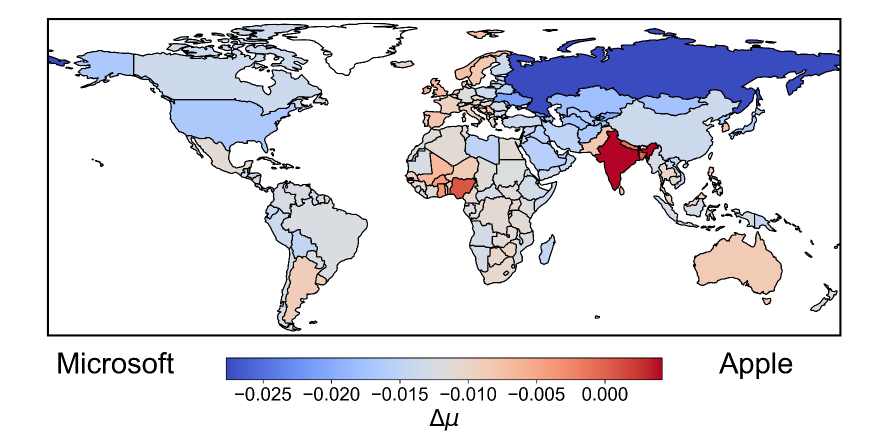}
\caption{\label{fig4}
Opinion polarization of world countries
for {\it Apple Inc.} ($\Delta \mu >0$) vs.
{\it Microsoft Corporation} ($\Delta \mu <0$), $\mu_0=-0.076$ following OPS 
for EN Wiki2025.
}
\end{center}
\end{figure}

In Figure~\ref{fig4}, we present the opinion polarization of world countries
with respect to two companies {\it Microsoft} (blue $\sigma=\mu=-1$)
vs. {\it Apple Inc.} (red $\sigma=\mu=1$) for the OPS case of the EN Wiki2025. 
The global opinion polarization $\mu_0=-0.076$ (average of $\sigma_i$ 
over all $N$ nodes and and all $N_r=10^5$ pathway realizations) 
is in favor of {\it Microsoft}.
The top countries with highest preferences $\Delta \mu > 0$ for
Apple Inc. are {\it India, Nigeria, Bangladesh, Nepal} 
and those with preferences $\Delta \mu < 0$ for Microsoft are 
{\it Russia, Georgia, Ukraine, Kyrgyzstan}. 
USA has a slight preference for
{\it Microsoft} and China has a slighter one also for {\it Microsoft}.
We attribute the significant negative $\Delta \mu$ value of Russia
to the fact that at the disappearance of the USSR
the Russian personal computer market was dominated by Microsoft
and this influence remained till recent times,
Apple computers were too expensive at those times.
A significant preference of India for {\it Apple Inc.}
is related to the presence of several direct
links pointing from Apple-related articles to India.

\begin{figure}[H]
\begin{center}
\includegraphics[width=0.7\textwidth]{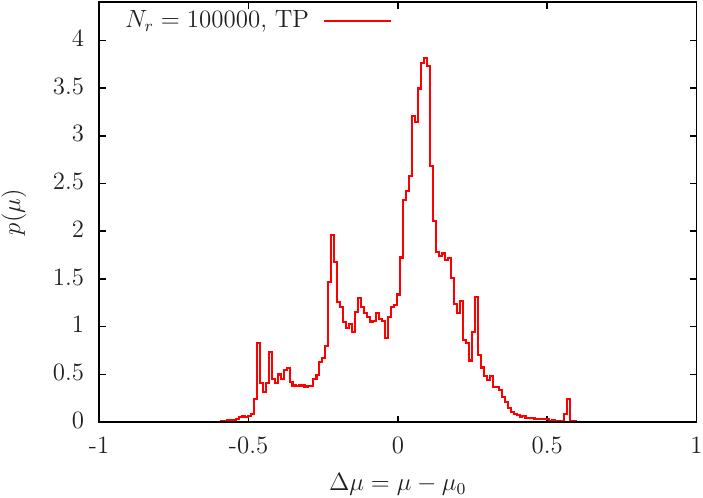}\\
\caption{\label{fig5}
  Probability density $p(\mu)$
for the OPS case in Eq.~(\ref{eqz}) for two fixed nodes with 
{\it Vladimir Putin} (red)
vs. {\it Donald Trump} (blue) for EN Wiki2025 (TP competition) 
with same bin number and normalization as in 
Figure~\ref{fig2}.
Here only one red curve of $p(\mu)$ versus $\Delta\mu=\mu-\mu_0$ for $N_r=100000$ is shown,
$\mu_0=0.11262$. 
}
\end{center}
\end{figure}

Another example is the competition between two pharmacological companies 
{\it Pfizer} (red $\mu=1$) vs. {\it Johnson \& Johnson} (blue $\mu=-1$) is 
presented in Appendix Figure~\ref{figA1} 
showing the polarization of countries on the world map.
The obtained results show a stronger influence of Pfizer with $\mu_0=0.049$
with its dominance in Canada, India, China even if Johnson \& Johnson 
is richer but Pfizer
is more ancient and more influential in WIN.

\subsection{Contest Donald Trump vs. Vladimir Putin}

Another example is a contest between
two groups with one red fixed node {\it Vladimir Putin}
and one fixed blue node {\it Donald Trump} for OPS case at EN Wiki2025. 

\begin{table}[H]
\centering
\begin{tabular}{llr}
\hline
Country & Article	&	$\Delta\mu$ \tiny{(Putin - Trump)}\\
\hline 
Argentina	&	Javier Milei	&	0.071	\\
Australia	&	Anthony Albanese	&	0.048	\\
Brazil	&	Luiz Inácio Lula da Silva	&	0.066	\\
Canada	&	Mark Carney	&	-0.042	\\
China	&	Xi Jinping	&	0.120	\\
France	&	Emmanuel Macron	&	0.120	\\
Germany	&	Friedrich Merz	&	0.200	\\
India	&	Narendra Modi	&	0.165	\\
Indonesia	&	Prabowo Subianto	&	0.161	\\
Italy	&	Giorgia Meloni	&	0.166	\\
Japan	&	Shigeru Ishiba	&	0.124	\\
Mexico	&	Claudia Sheinbaum	&	-0.040	\\
Russia	&	Vladimir Putin	&	0.887	\\
Saudi Arabia	&	Salman of Saudi Arabia	&	0.164	\\
South Africa	&	Cyril Ramaphosa	&	0.111	\\
South Korea	&	Lee Jae-myung	&	0.076	\\
Turkey	&	Recep Tayyip Erdoğan	&	0.234	\\
United Kingdom	&	Keir Starmer	&	0.082	\\
United States	&	Donald Trump	&	-1.113	\\
European Union	&	Ursula von der Leyen	&	0.155	\\
African Union	&	Mahamoud Ali Youssouf	&	0.176	\\
\hline 
\end{tabular}
\caption{Opinion polarization 
expressed by $\Delta\mu$ (following OPS for EN Wiki 2025), for leaders of G20. 
{\it Vladimir Putin} corresponds to $\mu=1$ and {\it Donald Trump}
corresponds to $\mu=-1$ with $\mu_0=0.113$. 
}
\label{table1}
\end{table}

The probability density distribution $p(\mu)$ for this contest in shown 
in Figure~\ref{fig5} with $\mu_0 = 0.113$ being in the favor of Putin. 
This corresponds to $f_p = (1+\mu_0)/2 = 0.5565$ votes for Putin and 
$f_b = (1-\mu_0)/2 = 0.4435$ votes for Trump
over all  nodes of  EN Wiki2025 network (white nodes are excluded).
The highest peak in Figure~\ref{fig5} at
$\mu \approx \mu_0 + 0.1$ corresponds to such articles as
{\it  Breton language,  French Polynesia,  Field (mathematics),
  Moses,  Mao Zedong,  Nairobi, Table tennis}.

In Table~\ref{table1} we show the opinion polarization 
$\Delta \mu_{G20} = \mu_{G20} - \mu_0$
for the contest Putin vs Trump for $G20$ political leaders representing
most influential world countries (including representatives of
European Union and African Union). It is surprising to see
that only leaders of Canada and Mexico have a polarization $\Delta \mu$ 
in favor of Trump
while all others have polarization $\Delta \mu$ in favor of Putin
with highest $\Delta \mu$ values for Turkey and Germany.

Here we should note that polarization in favor of one or another entry,
as the case Trump vs Putin, does not mean that a given entry
is favorable to Putin ($\Delta \mu >0$) or Trump ($\Delta \mu < 0$).
Indeed, it is difficult to think that Macron, Merz or Starmer
are favorable to Putin. The polarization $\mu$
measures the strength of links between entries or articles
but it does not take into account if a link has positive (like) or negative
(dislike) attitude. Thus it is more correct to interpret articles with a
high positive $\Delta \mu > 0$ value as strongly linked with or influenced by
Putin, and high negative $\Delta \mu < 0$ as those strongly linked with Trump.
Indeed, Canada and Mexico are strongly linked/influenced with/by USA and 
thus with Trump while Germany and Turkey are strongly linked/influenced 
with/by Russia and thus with Putin.
The information if links hold like/dislike (positive/negative) attitude is not  
accessible by the present network construction and INOF approach.

\begin{figure}[H]
\begin{center}
\includegraphics[width=0.65\textwidth]{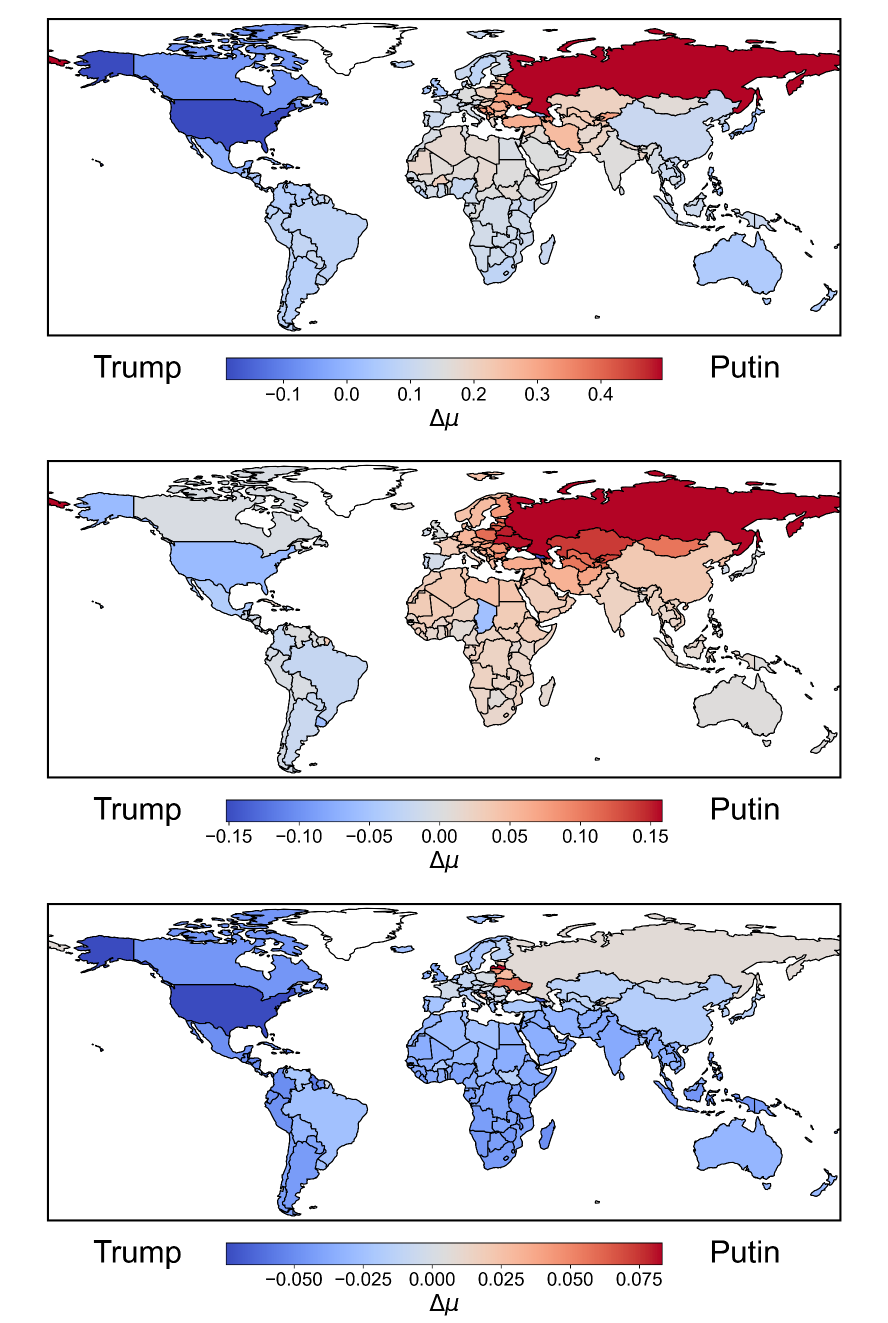}
\caption{\label{fig6}
Opinion polarization of world countries
for {\it Vladimir Putin } ($\Delta \mu >0$) vs.
{\it Donald Trump} ($\Delta \mu <0$) following OPS for EN  Wiki2025 (top panel with $\mu_0=0.113$), 
RU  Wiki2025 (middle panel with $\mu_0=0.592$) and ZH  Wiki2025 (bottom panel with $\mu_0=-0.340$).
}
\end{center}
\end{figure}

The influence of Trump vs Putin for the world countries
(represented by the Wikipedia articles of these countries)
is presented via the world map in Figure~\ref{fig6} with $\Delta \mu$
values of all countries and for three editions
EN, RU (for Russian), ZH (for Chinese) of Wikipedia 2025.
For EN Wiki2025  the positive polarization $\Delta \mu$ for
Putin extends to Russia, former republics of USSR, with strong influence
for Turkey, Iran, Bosnia and Herzegovina, Serbia, Albania.
The country polarization in favor of Trump includes
USA, Canada, Mexico, Latin America, UK, Australia, South Africa, Japan,
Ireland, New Zealand.
From Figure~\ref{fig6} middle panel for RU Wiki2025 
it follows that the polarization in favor of Putin
is significantly increased propagating to higher number of
countries including West Europe, Northern Africa,
slightly positive polarization of China.
Polarization in favor of Trump extends from USA to
Mexico, Brazil, Argentina, Chad. 
This is rather natural since the Russian Wikipedia edition 
favors the importance of Russia and Putin.
The polarization from ZH Wiki2025 
has almost all countries polarized in favor of Trump;
Russia has zero polarization $\Delta \mu \approx 0$.
Positive polarization for Putin
exists only in Latvia, Ukraine, Lithuania, Belarus, Bosnia and Herzegovina, 
Estonia, Moldova and Luxembourg.
This corresponds to the fact that events of special military
operation of Russia in Ukraine are strongly
linked with Putin explaining his significant
influence on Ukraine. 
However, it is somewhat surprising
that the ZH edition shows so strong polarization for Trump
with $\mu_0= -0.340$. We attribute this
to very strong commercial exchange
between China and USA.
Also one should take into account that
mainland China has its one analog of Wikipedia
in Chinese known as Baidu Baike.
Also the Chinese government has cut off access 
to the Chinese Wikipedia for residents of mainland China since 2019. 
Thus the contributions to the ZH edition at Wikipedia 
are coming mainly from outside of the mainland China 
and thus they may be rather different 
from the view points of the majority of China's population. 

\subsection{Competition Emmanuel Macron vs.  Marine Le Pen }

We also consider the INOF competition between
Emmanuel Macron ($\mu=1$) vs. Marine Le Pen ($\mu=-1$) with $\mu_0 = - 0.028$
from FR Wiki2024. These corresponds to rather close vote fractions
with $f_r =(1+\mu_0)/2 =0.4860$ for Macron and $f_b=(1-\mu_0)/2 = 0.5140$
for Le Pen (number of pathway realizations $N_r=10^6$).
The polarization opinions of 14 French political figures and
top 10 French richest persons from the Forbes list 2015-2024 are 
presented in Appendix Table~\ref{tabA1}. The results for these 
24 persons show that
only two of them have a polarization in favor of
Emmanuel Macron and other 22 in favor of Marine Le Pen.
At the same time the values of $\Delta \mu$ are mainly located at relatively
small values $|\Delta \mu| \sim 0.01$ being approximately
by a factor 10 smaller of values of for the contest of Trump-Putin
shown in Table~\ref{table1}. Due to these smaller $\Delta\mu$ values we
also compute an additional data set with $N_r=10^6$ to reduce the 
statistical errors (note that the FR Wiki2024 network is roughly 
2 times smaller in number of nodes and links  as compared to those of EN Wiki2025 thus
reducing the numerical effort). 
Table~\ref{tabA1} shows two columns of $\Delta\mu$ computed for $N_r=10^5$ 
and $N_r=10^6$ (the smaller data set is statistically 
independent and not included in the larger date set). The values are rather 
close and the small differences indicate the size of the typical fluctuations 
at $N_r=10^5$. 

The theoretical error of the $\mu_i$ values is 
$(1-\mu_i^2)/\sqrt{N_r}\approx 0.003$ (for $N=10^5$) or 
$\approx 0.001$ (for $N=10^6$). 
However, the more precise error estimation for $\Delta\mu_i=\mu_i-\mu_0$ using 
a subdivision of the data in 100 samples (described above after the 
discussion for Figure~\ref{fig2}) provides typical statistical errors 
of $\Delta\mu_i$ for the entries in Table~\ref{tabA1} which are 
roughly 3 times smaller than the theoretical error, i.e. $\approx 0.001$ 
(for $N=10^5$) or $\approx 0.0003$ (for $N_r=10^6$) which is 
due to rather strong correlations between $\mu_i$ and $\mu_0$. 
Therefore, despite the typical small values of $\Delta\mu_i\approx 0.01$ 
in Table~\ref{tabA1} 
their relative error is mostly only 3\% (for the $N_r=10^6$ data). 
Note that the error $0.0023$ of $\mu_0$ itself is closer to the 
theoretical error $0.003$ 
which is confirmed by the value $\mu_0=-0.025$ for $N_r=10^5$ while 
$\mu_0=-0.028$ for $N_r=10^6$.

For the world countries only a few of them have 
a favorable opinion polarization for Emmanuel Macron 
(e.g. {\it Andorra, Ivory Coast, San Marino, Finland,  Qatar})
while all others have $\Delta \mu < 0$ in favor of Marine Le Pen
with typical values $\Delta \mu \sim -0.01$.
This result is surprising since Emmanuel Macron, as a president,
has much more activity on the international level as compared 
to Marine Le Pen. Our interpretation is similar to those
discussed for the Trump-Putin case in Table~\ref{table1}:
a high opinion polarization in a favor of an entry
does not necessary mean positive or negative attitude to this entry
from the viewpoint of a given Wikipedia article but
shows that this entry produces a significant influence on this article
(positive or negative).

\section{Results for competition of three groups of entries}

It is possible to generalize the competition between 2 groups
to a competition between 3 groups. 
A similar case for a competition of 3 currencies in the world trade
has been considered in \cite{brics2024}. However, in \cite{brics2024}
there were no white nodes in the initial distribution
of nodes and the network size was very small 
representing only about 200 countries (nodes).

As in the case of competition of 3 currencies in \cite{brics2024}
the competition of 3 entries allows to highlight more
complex interactions between 3 selected entries
compared to the case of only 2 entries.

\begin{figure}[H]
\begin{center}
\includegraphics[width=0.9\textwidth]{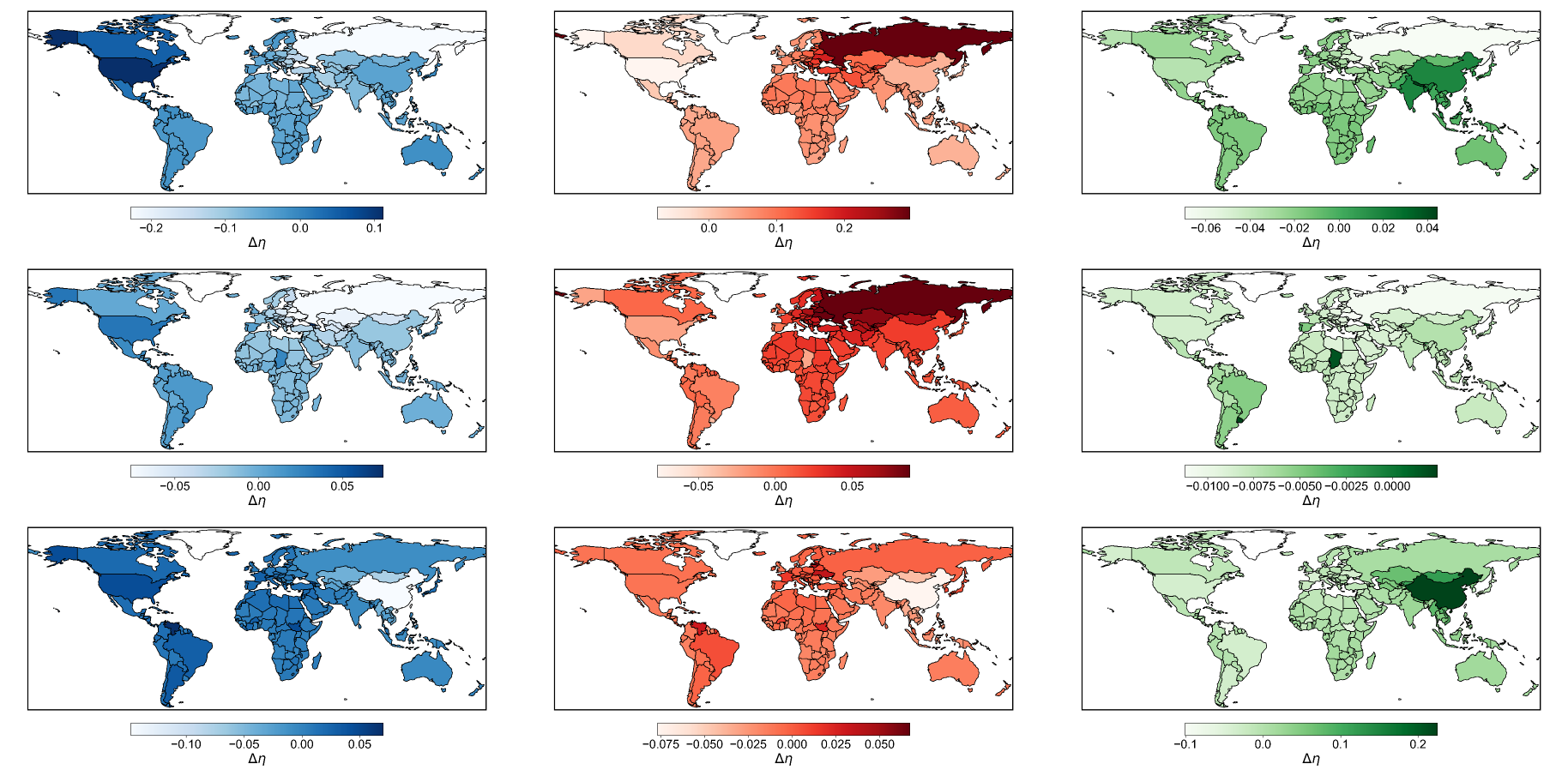}
  \caption{\label{fig7}
Opinion polarization of world countries
for {\it Donald Trump}, {\it Vladimir Putin }  and {\it Xi Jinping}
following OPS for EN  Wiki2025 (top panels),
RU  Wiki2025 (center row panels) and ZH  Wiki2025 (bottom panels).
Column panels show the cases of  {\it Donald Trump} in blue color on the left,
{\it Vladimir Putin } in red color in the center and
{\it Xi Jinping} in green color on the right.
The corresponding values of $\eta_{0,Trump}$,
$\eta_{0,Putin}$ and $\eta_{0,Jinping}$ are:
$\eta_{0,Trump}=0.398$, 
$\eta_{0,Putin} =0.4597$, 
$\eta_{0,Jinping} =0.1424$   for EN;
$\eta_{0,Trump}=0.193$, 
$\eta_{0,Putin} =0.779$, 
$\eta_{0,Jinping}=0.0282$ for RU; 
and 
$\eta_{0,Trump}=0.223$, $\eta_{0,Putin}=0.104$, 
$\eta_{0,Jinping}=0.673$. 
}
\end{center}
\end{figure} 

For the case of 3 competing groups, we compute 
for a given node $i$ three scores $Z_i(C)$ for three 
color values $C$ by:
\begin{equation}
	Z_i (C)= \sum_{j \neq i} \sigma_{j}(C) V_{ij} .
	\label{eqz3}
\end{equation}
Here $\sigma_j(C) = 1$ if node $j$ has color $C$ otherwise $\sigma_j(C)=0$ 
and in the computation of $Z_i(C)$ only nodes $i$ with color $C$ contribute. 
Note that the white color counts as an effective fourth color which 
has also a score but this fourth white score is not used in 
the spin update process of the Monte Carlo procedure. 
If among the three score values $Z_i(C)$ (for the three non-white colors) 
there is a single clear maximum color $C_{\rm max}$ 
with $Z_i(C_{\rm max})>Z_i(C)$ for $C_{\rm max}\neq C$, the node $i$ 
will acquire the new color $C_{\rm max}$. If there is no clear maximum, 
i.e. with at least two maximal identical 
values $Z_i(C_1)=Z_i(C_2)\ge Z_i(C_3)$ (for 
$C_1\neq C_2\neq C_3\neq C_1$) the color of node $i$ will not be changed 
(it may also stay white if it was white before). 
Note that we consider the three group competitions only for the OPS case 
with $V_{ij}=\tilde S_{ij}$ having fractional values. Therefore 
the scenario of two equal maximal scores and a third strictly smaller score 
($Z_i(C_3)<Z_i(C_{1,2})$) is very rare. However, the scenario of having 
three identical values being zero $Z_i(C_1)=Z_i(C_2)=Z_i(C_3)=0$ may happen 
quite regularly if all nodes $j$ with non-zero values of $V_{ij}$ in 
the sum (\ref{eqz3}) have still their initial blank color. 

After the color update of node $i$ the Monte Carlo procedure
is performed in the same way as for the above case (\ref{eqz}) of two colors:
the small number of nodes of the three groups with initially fixed color are 
never updated (they have a ``frozen''color) and the update procedure is 
done in random order for all other non-fixed nodes which have the 
white color as initial condition. 
A full update run is repeated up to 
$\tau=20$ iterations at which nearly all node color values are stable and in 
a steady state distribution. Finally, this procedure is repeated 
with the same initial condition but for $N_r$ different 
random pathway realizations in the update order 
which allows to compute averages 
and distributions of the obtained network color fractions. 

As for the two group competition, once a node switches from white to 
another color it cannot go back to the initial white color but even with this 
there is still a significant fraction of nodes which stay (nearly) 
always white for all $N_r$ pathway realizations. The sets of 
of ``white'' nodes essentially only depend on  the used Wikipedia edition (and 
not on the selected fixed color nodes for the competition) and these sets are 
also the same as for the two group competition. 

Formally the competition of 3 colors is different
from the Ising case of two colors with spins up or down
but we still keep  notations INOF, WIN for the case of 3 color competition
since it appeared originally from the Ising type spin relation (\ref{eqz})
with 2 colors. We note that in both procedures the spin/color information 
propagates from the initial groups with frozen spin/color through the 
network and after $\tau=20$ update iterations (per node) essentially 
all non-white nodes have a stable spin/color value which no longer changes. 
However, the final spin/color value of each node depends strongly on the 
selected random pathway for the update order (see also Figure~\ref{fig1}). 

We attribute to each of the three groups its own color $C\in\{R,G,B\}$ 
being red, green or blue (RGB) and compute for each (non-white) node 
the {\em color polarization} of color $C$ as the 
fraction $\eta_{C}(i)=n_C(i)/\sum_{C'\in\{R,G,B\}} n_{C'}(i)$ 
where $n_C(i)$ is the number of color $C$ outcomes of node $i$ 
in the $N_r$ pathway realizations. (Note 
that for the non-white nodes $i$ we typically have 
$\sum_{C'\in\{R,G,B\}} n_{C'}(i)\approx N_r$ while for the white nodes 
$j$ we have $\sum_{C'\in\{R,G,B\}} n_{C'}(j)\ll N_r$.) 

From $\eta_C(i)$ we compute for $C\in\{R,G,B\}$ 
its global network average (over non-white 
nodes) which is defined as $\eta_{0,C}$ and we characterize 
the node preference for color $C$ by the difference 
$\Delta\eta_C(i)=\eta_C(i)-\eta_{0,C}$ which represents the 
color preference of node $i$ in comparison to the global network 
color preference (both for color $C$). 

For the 3-entry analysis, where each of the 3 entries is mapped to 
an RGB color channel, we present two types of world map visualizations:
\textit{Monochromatic Maps:} Each map displays a single color 
    channel (Red, Green, or Blue). The color intensity is scaled by the 
    $\Delta\eta_C$ value, ranging from zero to maximum saturation.
    \textit{Multicolor Maps:} These maps use an RGB color triangle to represent the 
    combination of the 3 entries. The triangle is constructed such that each vertex 
    represents a pure color, corresponding to the maximum value of one component 
    ($\eta_R, \eta_G,$ or $\eta_B$) while the other two are at their minima.

Concerning the statistical error of $\eta_C(i)$ or $\Delta\eta_C(i)$, 
we mention that the theoretical error can be obtained in a similar way 
as for the case of two group competitions. Now, we use that $\eta_C(i)$ 
is the average (over the $N_r$ random pathway realizations) of 
$\sigma_i(C)$ a quantity which has values 0 or 1 such that 
$\sigma_i(C)^2=\sigma_i(C)$. This allows to compute the variance 
from its average and gives the theoretical 
error of $\eta_C(i)$ as $\sqrt{\eta_C(i)[1-\eta_C(i)]/(N_r-1)}
\approx 1/(2\sqrt{N_r})\approx 0.0015$ for $N_r=10^5$ and 
$\eta_C(i)\approx 0.5$ (value of maximal theoretical error) 
which is similar to the two color case. 
(The factor $1/2$ is a trivial effect of the formula 
$\eta_C(i)=(1\pm \mu)/2$ for the two color case.) We have also verified 
by the method of sample averages that the error of $\Delta\eta_i(C)$ 
is typically reduced by the same factor $2-3$ as for the two color case. 

In the following, we present results for the competition of 
three groups in next subsections for 3 types of groups 
being political leaders, countries and
society political concepts.

\subsection{Contest Trump, Putin, Xi Jinping}

In Figure~\ref{fig7} we present the world map of
(monochromatic) color polarization of countries
for {\it Donald Trump, Vladimir Putin, Xi Jinping}
from the view of the three Wiki2025 editions EN, RU, ZH.
The values of global color polarization 
$\eta_{0,Trump}, \eta_{0,Putin}, \eta_{0,Jinping}$
of these 3 political leaders are given
in the caption of Figure~\ref{fig7} (of course
$\eta_{0,Trump} + \eta_{0,Putin} + \eta_{0,Jinping} = 1$).

For EN Putin has the highest color
polarization being ahead of Trump and then Jinping.
For the pair Trump-Putin their relative polarization
remains approximately as in their own two group contest presented 
in the previous Section.
It is interesting to note that in this edition
there are more countries with $\Delta \eta$
in favor of Putin, also in this case
the maximal positive value $\Delta \eta(C) \approx 0.3$
is by a factor 3 higher than for the case of Trump
and by a factor 7 higher than for the case of Jinping. 

\begin{figure}[H]
\begin{center}
\includegraphics[width=0.65\textwidth]{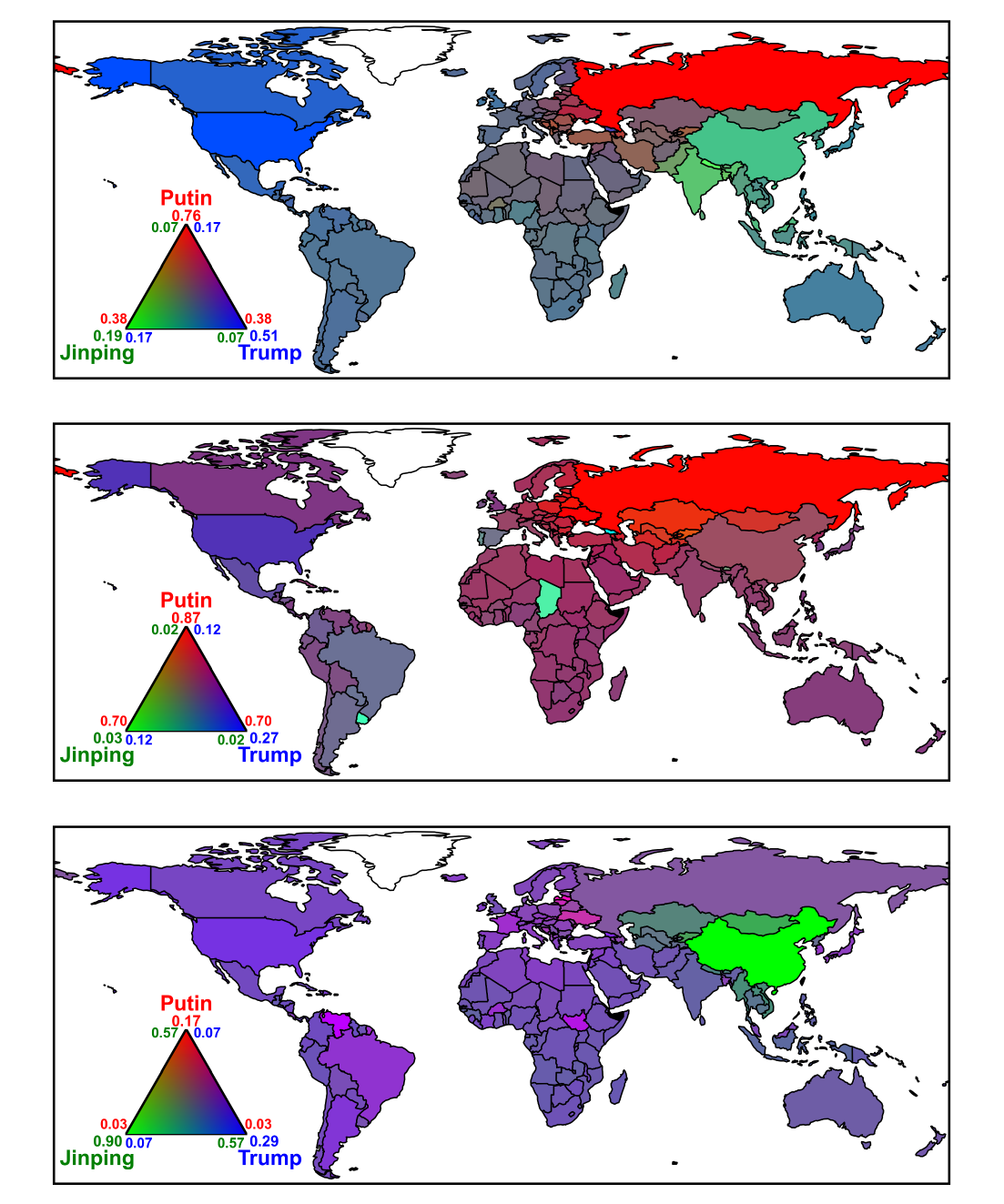}
  \caption{\label{fig8}
Opinion polarization of world countries $(\eta_{Trump},\eta_{Putin},\eta_{Jinping})$
for {\it Donald Trump}, {\it Vladimir Putin }  and {\it Xi Jinping}
following OPS for EN  Wiki2025  (top  panel), RU Wiki2025 (middle panel) and
ZH Wiki2025 (bottom panel) (same as Figure~\ref{fig7}).
The color mapping uses an \textbf{RGB color triangle} to visualize the 3-component polarization data 
($\eta_{\text{Putin}}, \eta_{\text{Jinping}}, \eta_{\text{Trump}}$). 
The vertices of the triangle are defined by the observed extrema in the data of countries. 
Specifically, the pure \textbf{Red vertex} represents the state where $\eta_{\text{Putin}}$ is 
at its maximum observed value ($\eta_{\text{Putin, max}}$) while both $\eta_{\text{Jinping}}$ and $\eta_{\text{Trump}}$ 
are at their minimums ($\eta_{\text{Jinping, min}}$, $\eta_{\text{Trump, min}}$). 
The \textbf{Green} and \textbf{Blue} vertices are defined analogously for Jinping and Trump, respectively. 
The resulting color for each country is an interpolation within this triangle,
where mixtures like magenta indicate 
high polarization towards both Putin (Red) and Trump (Blue),
and neutral colors represent a more balanced polarization.}
\end{center}
\end{figure} 

For RU the color polarization 
in favor of Putin extends even over more countries
than  expected. We attribute this to the fact that RU Wiki naturally gives
a higher preference to the president of Russia.

The ZH edition naturally places Jinping at the highest
global color polarization followed by
Trump and then Putin. The maximal polarization of countries
$\Delta \eta(C)$ is also by a factor 4 higher as compared 
to the cases of Trump and Putin
showing a high influence of Jinping of world  countries
from the view point of the ZH edition.

Any color can be presented as a combination of three colors red, green, 
blue (RGB).
Taking into account this property we can make a summation of 3-colors
world map in Figure~\ref{fig7} for each country using its corresponding
color average of $\eta_{Trump},\eta_{Putin}, \eta_{Jinping}$ 
(for each edition) 
and as a result to obtain a color RGB world map of countries. 
The result of this operation is presented in Figure~\ref{fig8}
for three editions EN, RU, ZH of Wiki 2025.

From the EN edition we see that the influence of Putin 
of course completely dominates in Russia
and also propagates to former USSR republics
(but it is not strong in the countries of Central Asia)
and few countries of East Europe
such as Romania, Hungary, Serbia and also
with a smaller strength to Turkey, Iran .
The influence of Trump naturally dominates USA and extends to  
Canada, Mexico, Latin America, UK, Australia, New Zealand, Japan.
South Korea. 
The influence of Xi Jinping from China extends to India, Pakistan and
countries of South East Asia.

In the case of the RU edition the influence of Putin
propagates from Russia to former Soviet republics,
Afghanistan and in a less strong way to Turkey, Iran,
Poland. The influence of Trump is restricted to USA extending to
Brazil, Argentina, Mexico. The clear influence of Xi Jinping
is well seen for Chad and Uruguay and is not well visible
for other countries that can be considered as
a significant exaggeration of the RU edition.
There are many countries with color being a mixture of red and blue
(between USA and Russia).

For the ZH edition the influence of Xi Jinping
propagates from China to Mongolia, Kazakhstan,
Kyrgyzstan as most obvious cases.
The country colors for influence of Trump and Putin
are mixed and give no clear preferences.

In  Appendix Figures~\ref{figA2},~\ref{figA3}
we show the density, or frequency, of articles
in the planes of color values $\eta_{Trump}, \eta_{Putin}, \eta_{Jinping}$
for the EN, RU, ZH editions that allows to see in a better
way the distribution of articles and their color polarizations. 
(more technical details in the captions of these figures). 
Note that for the case of a pure two group competition these type of figures 
would give straight lines on the antidiagonal (from $(0,1)$ to $(1,0)$), 
since e.g. 
$1=\eta_{Trump}+\eta_{Putin}$ for a the pure \textit{Trump-Putin} 
competition. 
In Appendix Figure~\ref{figA2}, the data are indeed somewhat 
concentrated close to this antidiagonal showing the that modifications 
due to the influence of the third group of \textit{Jinping} are rather 
modest for the editions of EN and RU.

\subsection{Contest USA, Russia, China}

\begin{figure}[H]
\begin{center}
\includegraphics[width=0.9\textwidth]{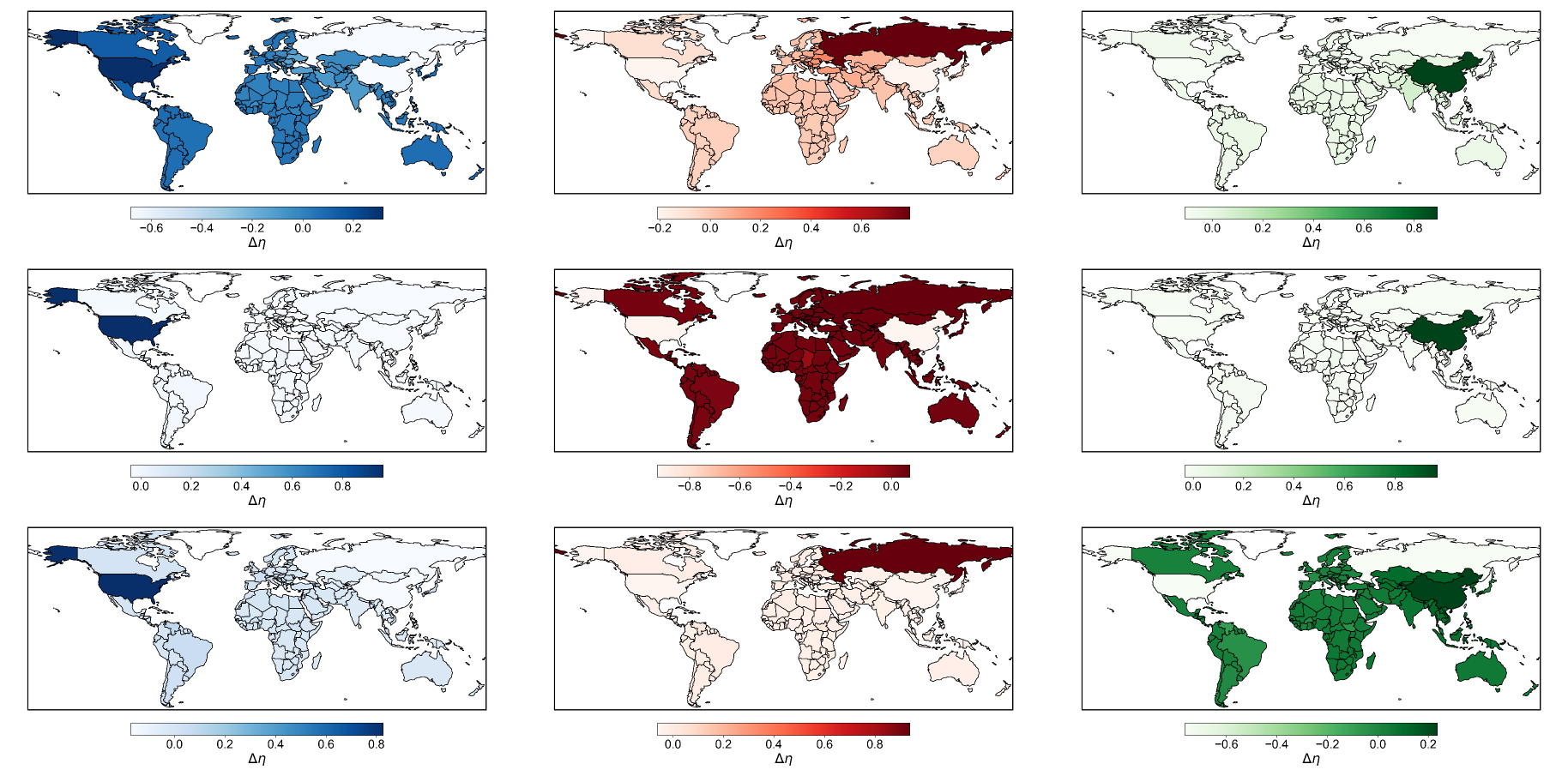}
  \caption{\label{fig9}
Opinion polarization of world countries
for {\it USA}, {\it Russia }  and {\it China} following OPS for EN  Wiki2025 (top panels),
RU  Wiki2025 (center row panels) and ZH  Wiki2025 (bottom panels).
Column panels show the case of  {\it  USA} in blue color on the left, {\it  Russia }
in red color in the center  and {\it China} in green color on the right.
The corresponding values of $\eta_{0,USA}$, $\eta_{0,Russia}$ and $\eta_{0,China}$ are:
$\eta_{0,USA}=0.682$, $\eta_{0,Russia} =0.209$, 
$\eta_{0,China} =0.109$   for EN;
$\eta_{0,USA}=0.040$, $\eta_{0,Russia} =0.927$, 
$\eta_{0,China}=0.033$ for RU; 
and 
$\eta_{0,USA}=0.174$, $\eta_{0,Russia}=0.062$, 
$\eta_{0,China}=0.764$.
}
\end{center}
\end{figure}

\begin{figure}[H]
\begin{center}
\includegraphics[width=0.65\textwidth]{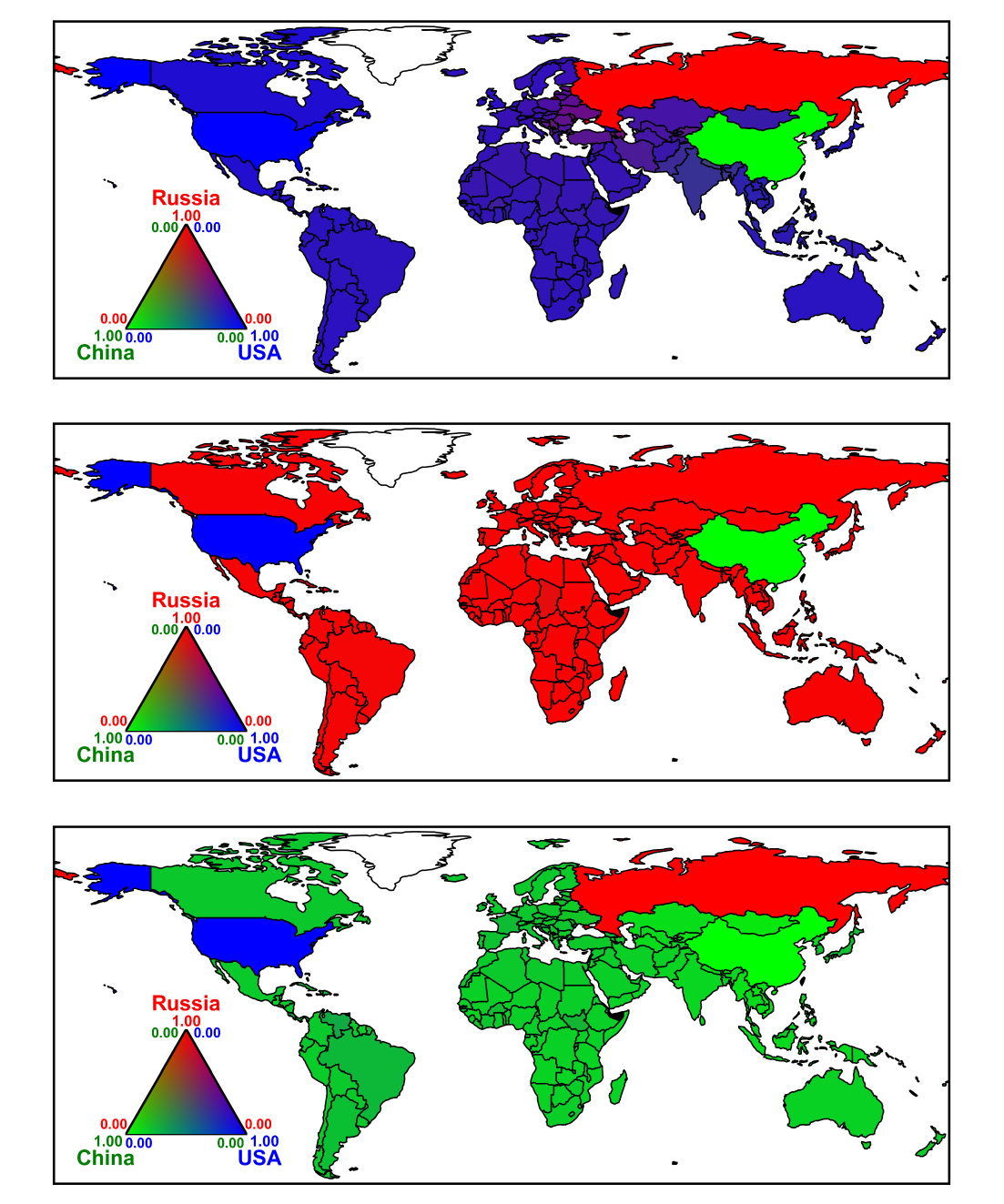}
  \caption{\label{fig10}
Opinion polarization of world countries $(\eta_{USA},\eta_{Russia},\eta_{China})$
for {\it USA}, {\it Russia }  and {\it China} following OPS for EN  Wiki2025  (top  panel),
RU Wiki2025 (middle panel) and ZH Wiki2025 (bottom panel) (same as 
Figure~\ref{fig8})
in color scale is given by normalized RGB value of Russia, China and USA respectively.}
\end{center}
\end{figure}

We present the results of contest between USA, Russia and China in
Figures~\ref{fig9},~\ref{fig10}. The presentation is similar
to the previous case of contest of
Donald Trump, Vladimir Putin, Xi Jinping
with the same three colors of opinion polarization. 
Here the situation is more standard 
with a dominance of USA in EN, Russia in RU, China in the ZH edition.
Each of the three editions seems to emphasize its color polarization
of other world countries in favor of their country of native language
(USA for EN, Russia for RU, China for ZH) as it is well visible in
corresponding panels of Figure~\ref{fig9}.
This effect is also clearly seen in Figure~\ref{fig10}
in the RGB representation.
Thus in its native edition
each of three countries considers that it has a dominant
influence on the main part of world countries.
In Appendix Figure~\ref{figA4} we show the
density distribution of articles (nodes)
in the planes of color opinion polarization 
($\eta_{USA}-\eta_{Russia}$-plane). Here in two out of three 
cases (EN and RU) the data is also somewhat close to the antidiagonal 
indicating a reduced influence of the third group for \textit{China}. 

It it interesting to compare the correlations
of polarization influence on 197 countries 
by political leaders: Donald Trump with those of USA,
Vladimir Putin with Russia and Xi Jinping with China.
To do this we present in Appendix Figure~\ref{figA5}
the plane of polarizations of 197 countries for editions EN, RU, ZH;
the values of Spearman correlation coefficient $\rho$ 
are given for each pair USA-Trump, Russia-Putin, China-Jinping
for these three editions. The two smallest correlator values are for the 
the pair China-Jinping (EN edition) with $\rho = 0.59$ 
and the pair USA-Trump (ZH edition) with $\rho=0.89$ 
while for all other pairs and editions we have $0.9 \leq \rho \leq  0.97$.
Such high correlator values clearly show the close link between the 
influence of political leaders on 197 world countries
to those of their own country influence on these 197 countries. 

\subsection{Contest Liberalism, Communism, Nationalism}

As a final example, we consider for the EN Wiki2025 edition the influence 
competition of the three social concepts 
{\it Liberalism} (blue), {\it Communism} (red) and {\it Nationalism} (green).

\begin{figure}[H]
\begin{center}
\includegraphics[width=0.9\textwidth]{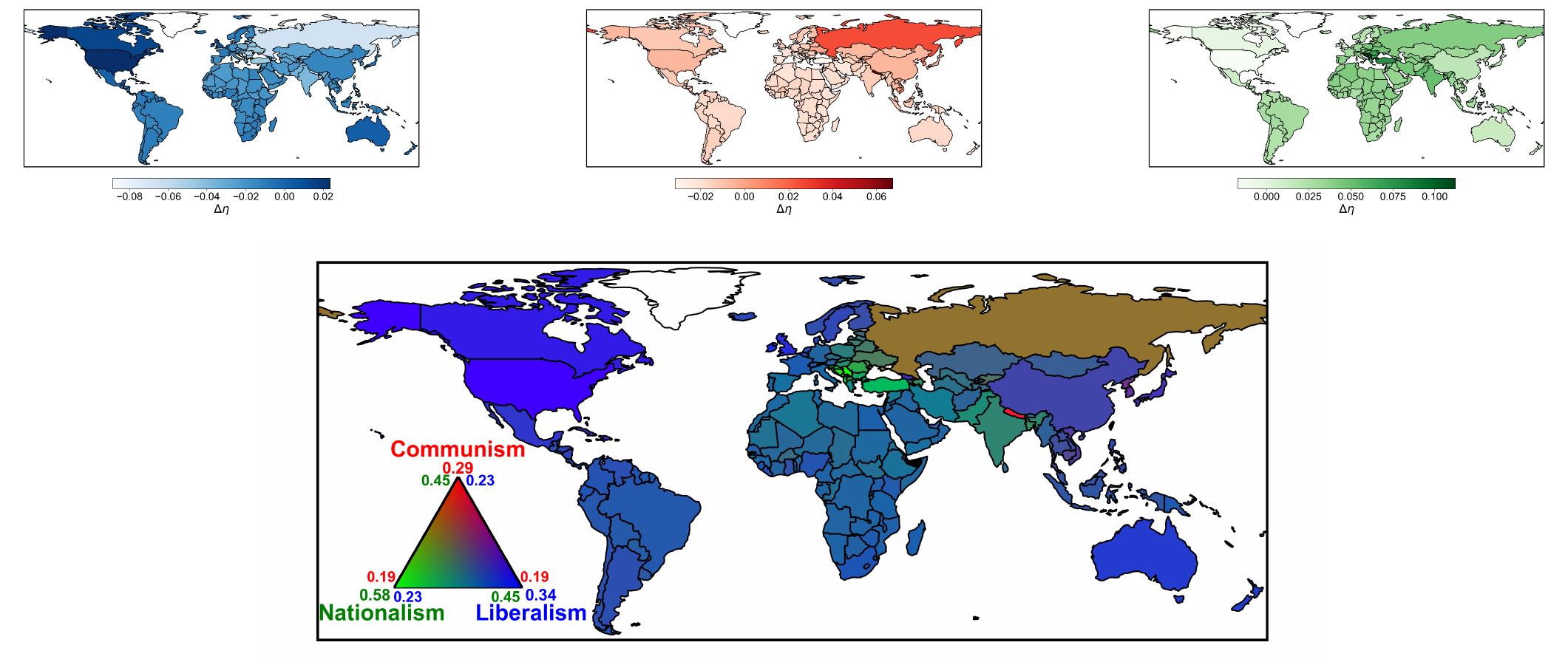}
  \caption{\label{fig11}
Opinion polarization of world countries
for {\it Liberalism}, {\it Communism }  and {\it Nationalism} following OPS
for EN  Wiki2025 (top panels). Column panels show the case of  {\it Liberalism}
in blue color on the left, {\it  Communism } in red color on the center  and {\it Nationalism}
in green color on the right.  In bottom panel the values of  $\eta_{Liberalism}$, $\eta_{Communism}$
and $\eta_{Nationalism}$ are shown in color scale given by normalized RGB values.
The corresponding values of $\eta_{0,Liberalism}$, $\eta_{0,Communism}$ and $\eta_{0,Nationalism}$ are:
$\eta_{0,Liberalism}=0.316$, $\eta_{0,Communism} =0.219$ and 
$\eta_{0,Nationalism} =0.466$ for EN.
}
\end{center}
\end{figure}

Their color influence are shown in Figure~\ref{fig11} 
    with 3 monochromatic color maps in the top 3 panels
    and an RGB map in the bottom panel.

    From the RGB map, we see that the strongest influence of Nationalism is for
    Serbia, Bosnia Herzegovina, Albania, Turkey, Romania, Bulgaria
    and in a less respect Pakistan, India, Bangladesh.
    The influence of Communism is rather local being represented only in Nepal.
    The influence of Liberalism is mostly spread over the world with
    the strongest adepts being UK, USA, Somaliland, Belgium, Netherlands 
    and many others.
    The color of Russia is at the middle between Communism and Nationalism.
    This distribution of influence of the three concepts corresponds to
    the political orientations of these countries. 
    Furthermore, in Appendix Figure~\ref{figA6}
    we present the histogram distribution 
    in $\eta_{Liberalism}-\eta_{communism}$-plane. 
    Here the articles are approximately distributed 
    around the coordinate $(0.33,0.23)$ corresponding to
    $\eta_{Nationalism} = 0.44$ which appears to be the typical
    polarization of these three concepts for 197 countries and 
    also all other Wikipedia articles. In particular, here the distribution 
    is not concentrated close to the antidiagonal as for certain other 
    cases, showing that none of the three groups has a significantly reduced 
    influence in comparison to the other two groups. 
    
\section{Effects of fluctuations at effective finite temperature}
For the competition of two groups with
different opinions (red vs, blue)
the relation (\ref{eqz}) for $Z_i$ determines
the condition of spin updates with $\sigma_i = 1$
if $Z_i  > 0$, $\sigma_i=-1$ if $Z_i <0$ 
and no spin change if $Z_i=0$. Such a condition
corresponds in the Monte Carlo process to the 
effective temperature $T=0$
since it gives a firm choice for the updated spin. 
It is interesting to analyze how stable this procedure is 
in presence of fluctuations
produced by a finite effective temperature $T$.
A finite $T$ value physically corresponds to
the presence of finite probabilities 
$W_\pm(i)$ ($W_+(i) + W_-(i)=1$) to obtain 
the new spin value $\sigma_i=\pm 1$. 

\begin{figure}[H]
\begin{center}
\includegraphics[width=0.75\textwidth]{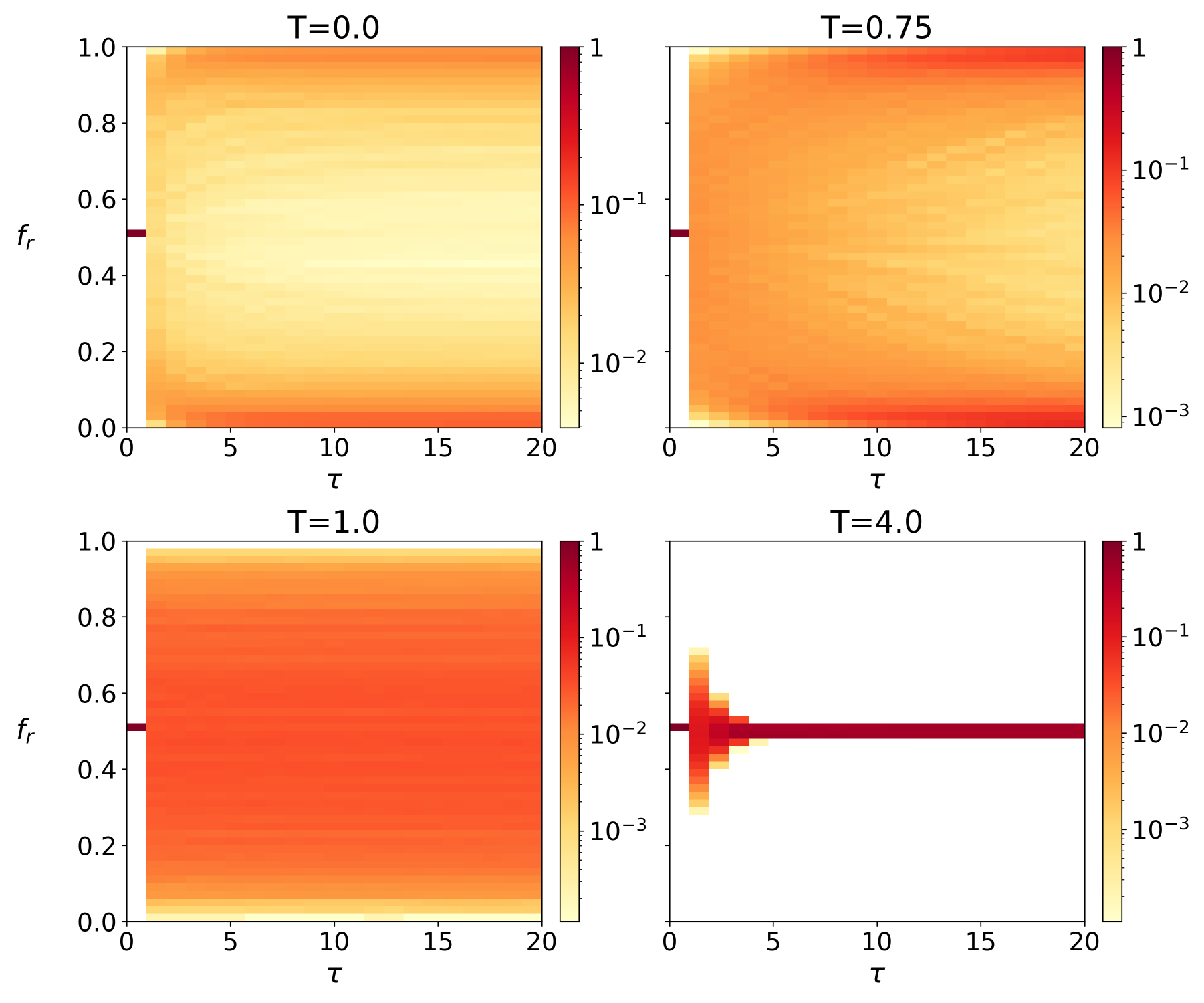}
  \caption{\label{fig12}
  Density of the fraction of red nodes, $f_r$, as a function of $\tau$, for the 
  competition between {\it Socialism, Communism} (red) 
  and {\it Capitalism, Imperialism} (blue), using the EN Wiki2024 dataset. 
  Each panel shows a density plot for $10^4$ pathway realizations 
on a logarithmic scale. 
  The distribution is normalized to one and uses 50 bins for 
the range $f_r \in [0, 1]$. 
  Panels correspond to different temperatures: $T=0$ (top left), $T=0.75$ (top right), 
  $T=1$ (bottom left), and $T=4$ (bottom right).
}
\end{center}
\end{figure} 

To model this situation
we write $Z_i=Z_+(i)-Z_-(i)$ as a difference of two positive 
quantities 
$Z_\pm(i)=\sum_{j\neq i,\,\sigma_j=\pm 1} V_{ij}\ge 0$ 
(i.e. sum only over all $j$ with either $\sigma_j=1$ or $\sigma_j=-1$ for 
the two cases $+$ or $-$ respectively). 
This is similar to the color score $Z_i(C)$ used 
in (\ref{eqz3}) if we use only two colors for spins $\pm 1$.  
Then the probabilities $W_\pm(i)$ are determined by the relations 
\begin{equation}
W_+(i) = \frac{{Z_+}^{\beta}(i)}{{Z_+}^{\beta}(i) +  {Z_-}^{\beta}(i)}
\quad,\quad
W_-(i) = \frac{{Z_-}^{\beta}(i)}{{Z_+}^{\beta}(i) +  {Z_-}^{\beta}(i)}
\quad,\quad T=\frac{1}{\beta}
\label{eqtemp}
\end{equation}
where during a Monte Carlo step the spin $i$
takes the value $\sigma_i=\pm 1$ with probability $W_{\pm} (i)$.
At $T=0$ ($\beta\to\infty$) we have $W_+(i)=1$ and $W_-(i)=0$ 
if $Z_i=Z_+(i)-Z_-(i)>0$ ($W_+(i)=0$ and $W_-(i)=1$ 
if $Z_i=Z_+(i)-Z_-(i)<0$) which reproduces 
the previous spin update condition based on $Z_i>0$ or $Z_i<0$. 
At high temperature $T \gg 1$ ($\beta\ll 1$) we 
have $W_+(i)\approx W_-(i)\approx 1/2$ such that the new spin 
value $\sigma_i=\pm 1$ is purely random with 
equal probabilities.

We mention that (\ref{eqtemp}) can be understood 
by introducing two virtual ``energy levels'' 
$\eps_\pm(i)=-\ln(Z_\pm(i))$ such ${Z_\pm(i)}^\beta=e^{-\beta \eps_\pm(i)}$
(for each node $i$ there is a different two level system). 
In this case the probabilities (\ref{eqtemp}) are just 
the usual probabilities of the levels $\eps_\pm(i)$ in the canonical ensemble 
at temperature $T$ for this two level system: 
$W_\pm(i)=e^{-\beta \eps_\pm(i)}/(e^{-\beta \eps_+(i)}+e^{-\beta \eps_-(i)})$. 

\begin{figure}[H]
\begin{center}
\includegraphics[width=0.75\textwidth]{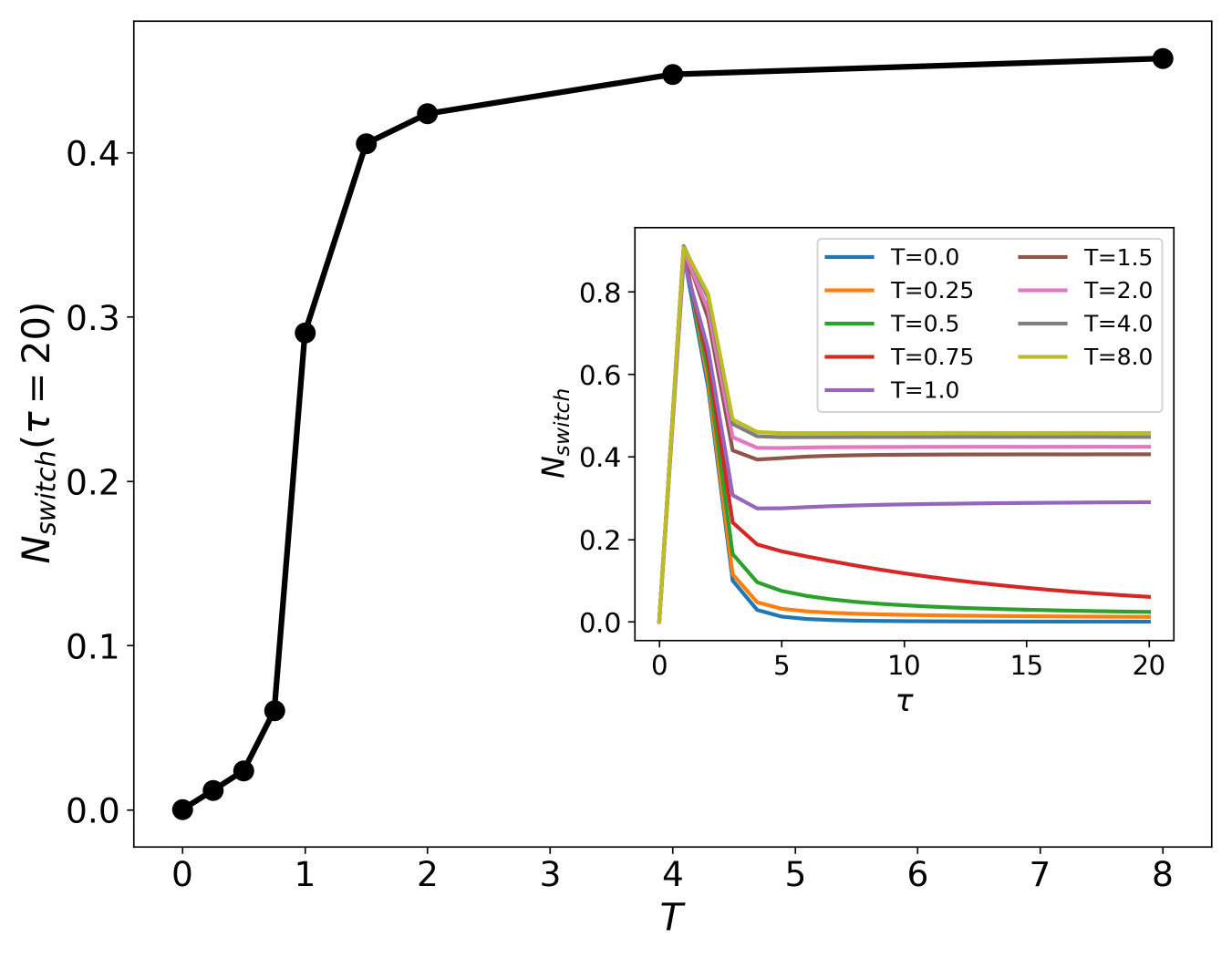}
  \caption{\label{fig13}  
Normalized number of switches, $N_{\text{switch}}$, between white, red, and blue nodes 
for the competition between {\it Socialism, Communism} (red) and {\it Capitalism, Imperialism} (blue), 
using the EN Wiki2024 dataset. The normalization is such that $N_{\text{switch}}=1$ corresponds 
to a number of switches equal to the total number of nodes. 
Main panel: $N_{\text{switch}}$ at maximal iteration time $\tau=20$ 
as a function of temperature 
$T$. Inset panel: The time evolution of $N_{\text{switch}}$ for several temperature values.
}
\end{center}
\end{figure}

The results for this finite temperature model of
fluctuations are shown in Figures~\ref{fig12} and \ref{fig13}.
Figure~\ref{fig12} shows that at $T\le 0.75$ 
the density distribution of the (network) fraction of red nodes $f_r$
in at maximal iteration time $\tau=20$ is concentrated 
to the two regions $f_r\approx 0$ and $f_r\approx 1$ (similarly 
as Figure~\ref{fig1} which corresponds to $T=0$). 
In contrast at $T=1$ this density has a broad homogeneous distribution
approximately in the range $ 0.2 \leq f_r \leq 0.8$
and at $T=4$ all density is located in a narrow range around $f_r=0.5$.
Indeed, at such a high temperature the probabilities
to have spin up or down from (\ref{eqtemp})
are very close $W_+ \approx W_- \approx 0.5$ 
and hence we have approximately half of spins up and half down.
The transition from a spin polarized steady-state at low temperatures
to a non-polarized one takes place in the vicinity of a 
certain critical temperature $T_c$.
Its value can be approximately determined by measuring the normalized number
of spin flips or spin switches $N_{switch}$ at \textit{maximal iteration time} 
$\tau=20$ as a function of temperature $T$. Up to now this quantity (at $T=0$) 
is essentially zero since at $\tau=20$ for a specific given pathway 
realization the spins of individual nodes are mostly in stable steady state. 

This dependence of $N_{switch}$ on temperature $T$ 
(and also on iteration time $\tau$) is shown in Figure~\ref{fig13}. 
This Figure shows that the critical temperature is $T_c \approx 1$
where we have a sharp increase of number of flips (at $\tau=20$) 
and a rapid growth of 
the normalized switch number $N_{switch}$. 
Thus the obtained results of Figures~\ref{fig12} and \ref{fig13}
show that the spin polarized phase remains stable for the temperature range
$0 \leq T \leq T_c \approx 1$ while above $T_c$ there is a melting of the 
polarized phase and we obtain a non-polarized liquid state at 
$T > T_c \approx 1$ at which individual spins no longer have 
stable values with respect to iteration time even at $\tau\ge 20$. 
In particular, we see that for a \textit{specific given random pathway realization} 
at $T=0$ or $T\ll T_c$ the spin values of individual nodes 
become stable in time (fluctuations discussed in the previous 
section are entirely due to the many \textit{different} random 
pathway realizations which produce \textit{different} steady states) 
while at $T\ge T_c$ there is no real spin-steady 
state (for a given pathway realization) and spins continue to be flipped 
even at $\tau\ge 20$. 

We argue that the main result of this effective 
temperature model (\ref{eqtemp}) is the fact that the polarized phase 
remains stable with respect to fluctuations at small or modest temperatures. 

\section{Discussion and conclusion}

In this work we described the process of opinion formation
appearing in Wikipedia Ising Networks (WIN) being
based on an asynchronous Monte Carlo procedure.
This INOF approach is determined by a simple natural rule
that the opinion of a given node (article, user)
in a network is determined by a majority
opinion of other nodes connected to 
this given node. We discussed two
possible voting procedures: OPA case where 
vote contributions 
are given by a sum over elements of the adjacency
matrix $A_{ij}$ going to a selected given node $i$ 
or the OPS case when the the weight of a vote
is given by an element of
normalized matrix of Markov transitions
$S_{ij}$ (see \ref{eqz}).
Only the OPA case was considered in previous studies \cite{inof24}.
We show that these two vote options give similar
results but specific vote polarizations
may be different. We think that
both vote options OPA and OPS can be suitable
for the description of opinion formation
on networks. Thus for the protein-protein
interaction networks we think that
the OPS case, used for the MetaCore protein network in
\cite{fibrosis2}, is more correct since
the interaction capacity of a given protein
is bounded by various chemical processes.
The important new element of the INOF approach
is the presence of white nodes
with undefined opinion at the initial stage of the
asynchronous Monte Carlo process.
Our results show that the spin polarized steady-state
remains stable with respect to small fluctuations at
an effective temperature below a certain critical border
while above this border there is a melting of this phase
and a transition to a liquid non-polarized spin phase. 

We also demonstrated that the situation
with competition between two groups
with fixed red/blue opinions
can be generalized to the case of three
competing groups with fixed opinions (red, green, blue)
and that in this case the generalized INOF approach
leads to fair results for WIN,

For the EN, RY, ZH editions of Wikipedia 2025
we compared opinions of different cultural views
of these editions with respect to
political leaders Donald Trump,
Vladimir Putin, Xi Jinping
and determined their influence on 197 world counties.
Surprisingly Putin happens to produce a higher
polarization influence in the EN edition.
With the INOF approach we also determined 
the influence of USA, Russia, China
on other countries for these 3 editions.
We also showed that other types of contests 
can be studied like the competition between
Liberalism, Communism, Nationalism.

The described INOF approach is generic and can be applied to various
directed networks. Thus in \cite{fibrosis2} this approach
allowed to describe myocardial fibrosis progression
in the MetaCore network of protein-protein interactions.
Also, a variation of this approach (without white nodes)
determines dominant features of trade currencies
in the World Trade Network from the UN COMTRADE database \cite{brics2024}.

Of course, the Wikipedia networks have important exceptional
features as compared to other networks:
the meaning of their nodes is very clear, 
they enclose all aspects of nature and human activity
and the presence of multiple language editions
allows to analyze various cultural views of humanity.
Thus we hope that the INOF approach to
Wikipedia Ising networks will find
diverse interesting applications.

\section*{Appendix}
\setcounter{equation}{0}
\renewcommand{\theequation}{A\arabic{equation}}
\setcounter{figure}{0}
\setcounter{section}{0}
\setcounter{table}{0}
\renewcommand\thefigure{A\arabic{figure}}
\renewcommand\thetable{A\arabic{table}}
\renewcommand\thesection{A\arabic{section}}
\renewcommand{\figurename}{Appendix Figure}
\renewcommand{\tablename}{Appendix Table}


Here we present additional Figures and one additional data table 
related to the main part of the article. Below, we mostly only give 
a short description of them and for a more detailed discussion of 
this materiel we refer to the main text. 

\begin{figure}[H]
\begin{center}
\includegraphics[width=0.65\textwidth]{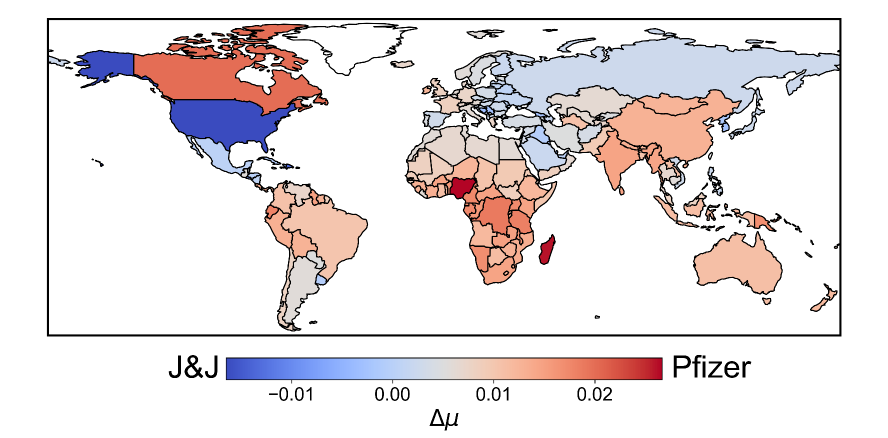}
  \caption{\label{figA1}
Opinion polarization of world countries
for {\it Pfizer} ($\Delta \mu >0$) vs.
{\it Johnson \& Johnson} ($\Delta \mu <0$), $\mu_0=0.049$ following OPS 
for EN Wiki2025.
}
\end{center}
\end{figure}

Appendix Figure~\ref{figA1} shows the opinion polarization of world countries
for {\it Pfizer} ($\Delta \mu >0$) vs. {\it Johnson \& Johnson} ($\Delta \mu <0$) 
using the EN Wiki2025 edition. Globally, the influence of Pfizer seems 
to be stronger, however Johnson \& Johnson are dominating the USA. 

\begin{table}[h]
\centering
\begin{tabular}{lrr}
\hline
Article	&	$\Delta\mu$ \tiny{(Macron - Le Pen)} $N_r=10^5$ & 
$\Delta\mu$ \tiny{(Macron - Le Pen)} $N_r=10^6$ \\
\hline
François Hollande	&	-0.0095	&	-0.0095	 \\
Jean-Luc Mélenchon	&	-0.0120	&	-0.0113	 \\
Édouard Philippe	&	-0.0104	&	-0.0099	 \\
Nicolas Sarkozy	&	-0.0098	&	-0.0099	 \\
Dominique Strauss-Kahn	&	-0.0098	&	-0.0101	 \\
Manuel Valls	&	-0.0103	&	-0.0103	 \\
Dominique de Villepin	&	-0.0099	&	-0.0100	 \\
Éric Zemmour	&	-0.0113	&	-0.0117	 \\
Gabriel Attal	&	-0.0105	&	-0.0102	 \\
François Bayrou	&	-0.0112	&	-0.0107	 \\
Éric Ciotti	&	-0.0141	&	-0.0135	 \\
Jordan Bardella	&	-0.0155	&	-0.0150	 \\
Rachida Dati	&	-0.0106	&	-0.0106	 \\
Bruno Retailleau	&	-0.0107	&	-0.0105	 \\
\hline
Bernard Arnault	&	-0.0078	&	-0.0077	 \\
Françoise Bettencourt Meyers	&	-0.0083	&	-0.0069	 \\
Alain Wertheimer	&	0.0289	&	0.0320	 \\
Gérard Wertheimer	&	0.0289	&	0.0320	 \\
François Pinault	&	-0.0075	&	-0.0074	 \\
Emmanuel Besnier	&	0.0172	&	0.0175	 \\
Nicolas Puech	&	-0.0006	&	-0.0004	 \\
Vincent Bolloré	&	-0.0104	&	-0.0095	 \\
Xavier Niel	&	-0.0088	&	-0.0084	 \\
Carrie Perrodo	&	-0.0019	&	-0.0005	 \\
\hline
\end{tabular}
\caption{Opinion polarization 
expressed by $\Delta\mu$ (following OPS for FR Wiki2024), for important 
personalities from French politics (top) and French richest persons
(following Forbes top 10 ranking 2015-2024) (bottom). 
{\it Emmanuel Macron} corresponds to $\mu=1$ and {\it Marine Le Pen}
corresponds to $\mu=-1$ with $\mu_0=-0.0252\pm 0.0023 $ for $N_r=10^5$ 
and $\mu_0=-0.0279\pm 0.0008$ for $N_r=10^6$. 
The 2nd (3rd) column corresponds to data for $N_r=10^5$ ($N_r=10^6$) 
with a typical statistical error $0.001$ ($0.0003$). 
The data for $N_r=10^5$ is statistically 
independent and not a subset of the data for $N_r=10^6$. 
}
\label{tabA1}
\end{table}

Appendix Table~\ref{tabA1}, provides the values of $\Delta\mu_i$ 
for certain French political or rich personalities for the competition 
{\it Emmanuel Macron} ($\mu=1$) vs. {\it Marine Le Pen} ($\mu=-1$) 
in FR Wiki2024. For most 
entries in this table there is a slight preference for Le Pen with 
typical values $\Delta\mu_i\approx -0.01$. Since these values are close 
to zero two data columns for $N_r=10^5$ and $N_r=10^6$ are shown which 
clarifies that the statistical fluctuations are typically well below 
$|\Delta\mu_i|$. 

\begin{figure}[H]
\begin{center}
\includegraphics[width=0.9\textwidth]{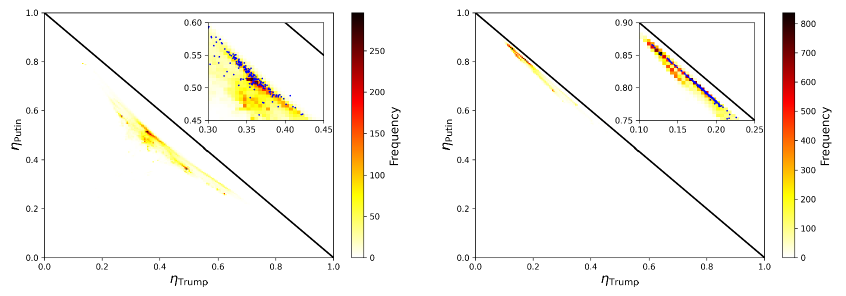}
  \caption{\label{figA2}
  Histogram of opinion polarization in the $(\eta_{Trump},\eta_{Putin})$
  plane for {\it Donald Trump}, {\it Vladimir Putin}, and {\it Xi Jinping} across all articles
  following OPS for EN Wiki2025 (left panel), RU Wiki2025 (right panel)).
  Note that for each article, the sum of polarization values is normalized to 1,
  and therefore $\eta_{Jinping}=1-\eta_{Trump}-\eta_{Putin}$.
  The corresponding insets show the same histogram zoomed in on
  the most populated region of the plane, with blue circles representing the country articles.}
\end{center}
\end{figure} 

\begin{figure}[H]
\begin{center}
\includegraphics[width=0.9\textwidth]{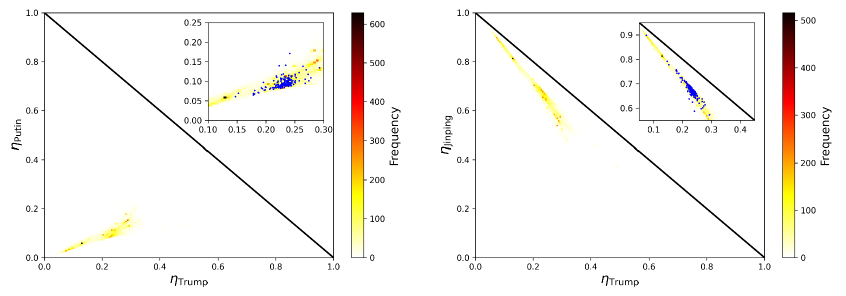}
  \caption{\label{figA3}
  Histogram of opinion polarization in the $(\eta_{Trump},\eta_{Putin})$ plane (left panel)
  and $(\eta_{Trump},\eta_{Jinping})$ plane (right panel)
  for {\it Donald Trump}, {\it Vladimir Putin}, and {\it Xi Jinping}
  across all articles following OPS case for ZH Wiki2025.
  Note that for each article, the sum of polarization values is normalized to 1,
  $\eta_{Jinping}+\eta_{Trump}+\eta_{Putin}=1$. The corresponding insets show
  the same histogram zoomed in on the most populated region of the plane,
  with blue circles representing the country articles.}
\end{center}
\end{figure} 

\begin{figure}[H]
\begin{center}
\includegraphics[width=0.9\textwidth]{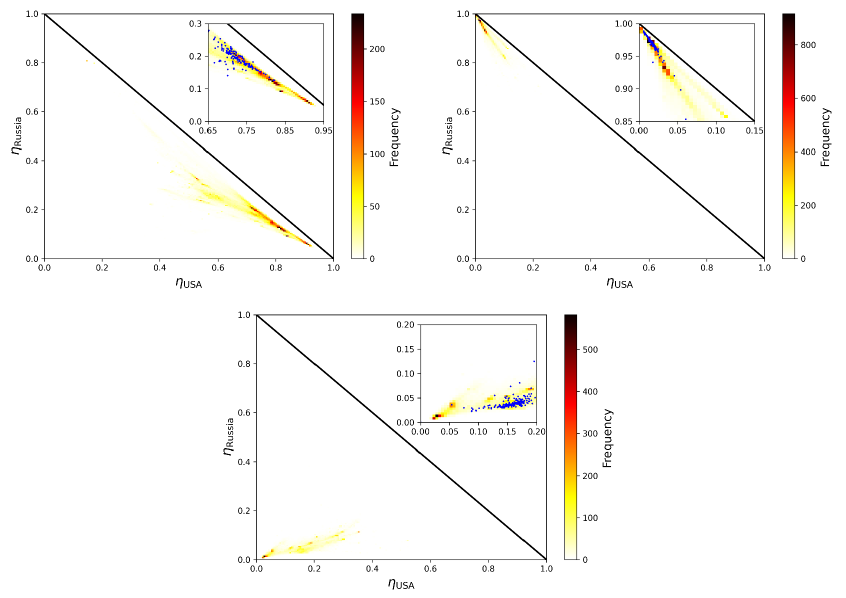}
  \caption{\label{figA4}
  Histogram of opinion polarization in the $(\eta_{USA},\eta_{Russia})$ plane for
  {\it USA}, {\it Russia}, and {\it China} across all articles following OPS
  for EN Wiki2025 (top left panel), RU Wiki2025 (top right panel), and ZH Wiki2025 (bottom panel).
  Note that for each article, the sum of polarization values is normalized to 1,
  and therefore $\eta_{China}=1-\eta_{USA}-\eta_{Russia}$.
  The corresponding insets show the same histogram zoomed in on the most populated region
  of the plane, with blue circles representing the country articles.}
\end{center}
\end{figure} 

Appendix Figures~\ref{figA2}-\ref{figA4} provide color plots of histogram 
distributions for certain three group competitions in the plane of 
two color polarization values $\eta_{C_1}-\eta_{C_2}$ 
(see main text for a detailed discussion). In certain cases the 
data is close to the antidiagonal (with $1\approx \eta_{C_1}+\eta_{C_2}$) 
indicating that typical values of the third color polarization 
$\eta_{C_3}=1-(\eta_{C_1}+\eta_{C_2})$ are rather small. 

\begin{figure}[H]
\begin{center}
\includegraphics[width=0.9\textwidth]{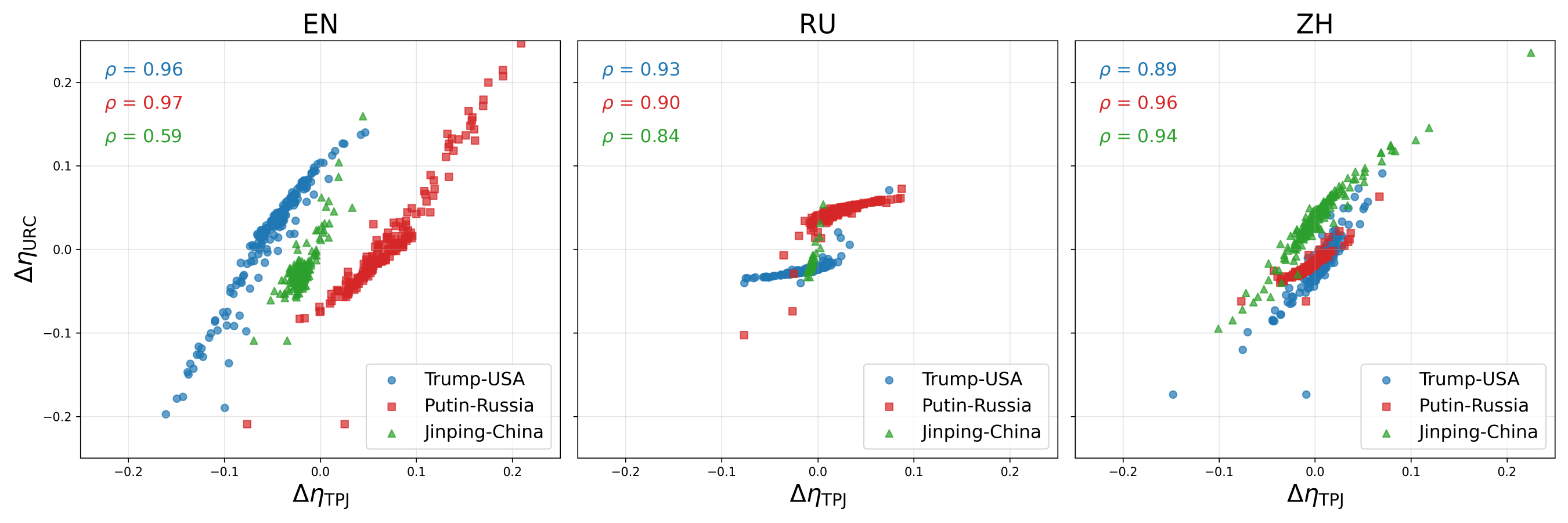}
  \caption{\label{figA5}
    Plane of color polarizations $\Delta\eta_{URC}$
    for the contest of USA, Russia, China of Figure~\ref{fig9} vs.
    those $\Delta \eta_{TPJ}$ for the contest of 3 political leaders of Figure~\ref{fig7}
    with data shown for all 197 countries.
The metrics are derived from the triads (USA, Russia, China) and (Trump, Putin, Jinping),
 respectively. Panels correspond to the datasets from Wiki 2025 for: (left) English (EN), 
 (center) Russian (RU), and (right) Chinese (ZH). 
 The Spearman correlation coefficient, $\rho$, is reported for each panel.
}
\end{center}
\end{figure} 

Appendix Figure~\ref{figA5} illustrates correlations 
between the three group competitions of \textit{USA, Russia, China} 
and \textit{Trump, Putin, Jinping}. 

\begin{figure}[H]
\begin{center}
\includegraphics[width=0.65\textwidth]{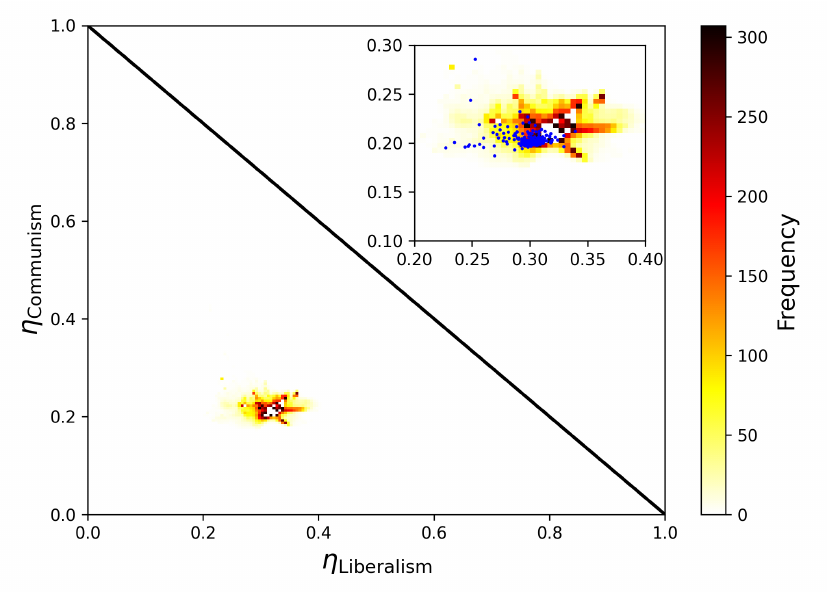}
  \caption{\label{figA6}
  Histogram of opinion polarization in the $(\eta_{Liberalism},\eta_{Communism})$
  plane for {\it  Liberalism}, {\it  Communism}, and {\it  Nationalism} across
  all articles following OPS for EN Wiki2025. Note that for each article, the sum of
  polarization values is normalized to 1, and therefore $\eta_{Nationalism}=1-\eta_{Liberalism}-\eta_{Communism}$.
  The corresponding insets show the same histogram zoomed in on the most populated region of the plane,
  with blue circles representing the country articles.}
\end{center}
\end{figure} 

Appendix Figure~\ref{figA6} is similar to 
Appendix Figures~\ref{figA2}-\ref{figA4} for the case of 
the three group competition \textit{Liberalism, Communism, Nationalism} 
for EN Wiki2025.


\begin{acknowledgments}
The authors acknowledge support from the grant
 ANR France project
NANOX $N^\circ$ ANR-17-EURE-0009 in the framework of 
the Programme Investissements d'Avenir (project MTDINA).
\end{acknowledgments}







\begin{thebibliography}{99}

  \bibitem[{Wikipedia contributors}(2024)]{soc1}
{Wikipedia contributors}.
\newblock Social media use in politics --- {Wikipedia}{,} The Free Encyclopedia.
\newblock
  \url{https://en.wikipedia.org/wiki/Social_media_use_in_politics},
  2024.
\newblock [Online; accessed 12-March-2025].

\bibitem[Fujiwara et~al.(2023)]{soc2}
Fujiwara, T.; Muller, K.; Schwarz, C.
\newblock The Effect of Social Media on Elections: Evidence from The United States
\newblock {\em J. Eur. Economi Ass..} {\bf 2023}, jvad058.
\newblock {\url{https://doi.org/10.1093/jeea/jvad058}}.

\bibitem[Castellano et~al.(2009)Castellano, Fortunato, and Loreto]{fortunato09}
Castellano, C.; Fortunato, S.; Loreto, V.
\newblock Statistical physics of social dynamics.
\newblock {\em Rev. Mod. Phys.} {\bf 2009}, {\em 81},~591--646.
\newblock {\url{https://doi.org/10.1103/RevModPhys.81.591}}.

\bibitem[Dorogovtsev(2010)]{dorogovtsev10}
Dorogovtsev, S.
\newblock {\em Lectures in Complex Networks}; Oxford University Press,  2010.

\bibitem[Galam(1982)]{galam82}
Galam, S.; Gefen Y.; Shapi Y.
\newblock Sociophysics: A new approach of sociological collective behaviour.
     I. mean-behaviour description of a strike.
\newblock {\em Journal of Mathematical Sociology} {\bf 1982}, {\em 9(1)},~1--13.
\newblock {\url{https://www.tandfonline.com/doi/abs/10.1080/0022250X.1982.9989929}}.

\bibitem[Galam(1986)]{galam86}
Galam, S.
\newblock Majority rule, hierarchical structures, and democratic
  totalitarianism: A statistical approach.
\newblock {\em Journal of Mathematical Psychology} {\bf 1986}, {\em
  30},~426--434.
\newblock {\url{https://doi.org/https://doi.org/10.1016/0022-2496(86)90019-2}}.

\bibitem[Sznajd-Weron and Sznajd(2000)]{sznajd00}
Sznajd-Weron, K.; Sznajd, J.
\newblock Opinion evolution in closed community.
\newblock {\em International Journal of Modern Physics C} {\bf 2000}, {\em
  11},~1157--1165,
\newblock {\url{https://doi.org/10.1142/S0129183100000936}}.

\bibitem[Sood and Redner(2005)]{sood05}
Sood, V.; Redner, S.
\newblock Voter Model on Heterogeneous Graphs.
\newblock {\em Phys. Rev. Lett.} {\bf 2005}, {\em 94},~178701.
\newblock {\url{https://doi.org/10.1103/PhysRevLett.94.178701}}.

\bibitem[Watts and Dodds(2007)]{watts07}
Watts, D.J.; Dodds, P.S.
\newblock {Influentials, Networks, and Public Opinion Formation}.
\newblock {\em Journal of Consumer Research} {\bf 2007}, {\em 34},~441--458,
  \href{http://arxiv.org/abs/https://doi.org/10.1086/518527}{{\normalfont
  [https://doi.org/10.1086/518527]}}.
\newblock {\url{https://doi.org/10.1086/518527}}.

\bibitem[Galam(2008)]{galam08}
Galam, S.
\newblock Sociophysics: a review of Galam models.
\newblock {\em International Journal of Modern Physics C} {\bf 2008}, {\em
  19},~409--440,
\newblock {\url{https://doi.org/10.1142/S0129183108012297}}.

\bibitem[Schmittmann and Mukhopadhyay(2010)]{schmittmann10}
Schmittmann, B.; Mukhopadhyay, A.
\newblock Opinion formation on adaptive networks with intensive average degree.
\newblock {\em Phys. Rev. E} {\bf 2010}, {\em 82},~066104.
\newblock {\url{https://doi.org/10.1103/PhysRevE.82.066104}}.

%

\bibitem[Ermann and Shepelyansky(2024)]{inof24}
           Ermann, L.; Shepelyansky, D.L.
           \newblock {Confrontation of capitalism and socialism in Wikipedia Networks}.
           \newblock {Information } {\bf 2024}, {\em 15},~571
           \newblock {\url{https://doi.org/10.3390/info15090571}}.

\bibitem[Hopfield(1982)]{memory1}
Hopfield, J.J.
\newblock Neural networks and physical systems with
    emergent collective computational abilities.
\newblock {\em Proc. Nat. Acad. Sci.} {\bf 1982}, {\em 79(8)},~2554--2558.
\newblock {\url{https://doi.org/10.1073/pnas.79.8.2554}}.

\bibitem[Benedetti et~al.(2024)Benedetti, Carillo, Marinari and Mezard]{memory2}
Benedetti, M.; Carillo, L.; Marinari, E.; Mezard, N.
\newblock Eigenvector dreaming
\newblock {\em J. Stat. Mech.} {\bf 2024}, 013302.
\newblock {\url{https://doi.org/10.1088/1742-5468/ad138e}}.

\bibitem[Dorogovtsev et~al.(2002)Dorogovtsev, Goltsev and Mendes]{dorogovtsev}
Dorogovtsev, S.N.; Goltsev, A.V.;  Mendes, F.F.
\newblock Ising model on networks with an arbitrary distribution of connections.
\newblock {\em Phys. Rev. E} {\bf 2002}, {\em 66}, 016104.
\newblock {\url{https://doi.org/10.1103/PhysRevE.66.016104}}.

\bibitem[Bianconi(2002)]{bianconi}
Bianconi, G.
\newblock Mean field solution of the Ising model on a Barabási–Albert network.
\newblock {\em Phys. Lett. A} {\bf 2002}, {\em 303}, 166.
\newblock {\url{https://doi.org/10.1073/pnas.79.8.2554}}.

\bibitem[Frahm et al.(2024)]{fibrosis2}
  Frahm K.M., Kotelnikova E., Kunduzova O. and Shepelyansky D.L.
  \newblock Fibroblast-Specific Protein-Protein Interactions
  for Myocardial Fibrosis from MetaCore Network
\newblock {\em Biomolecules} {\bf 2024}, {\em 14},~1395.
\newblock {\url{https://doi.org/10.3390/biom14111395}}.

\bibitem[Coquide et~al.(2024)Coquide, Lages, and Shepelyansky]{brics2024}
Coquide, C.; Lages, J.; Shepelyansky, D.L.
\newblock Opinion Formation in the World Trade Network.
\newblock {\em Entropy} {\bf 2024}, {\em 25(2)},~141.
\newblock {\url{https://doi.org/10.3390/e26020141}}.


\bibitem[Reagle(2010)]{wikiacad1}
Reagle, J.M.
\newblock {\em Good faith collaboration: the culture of Wikipedia.}; MIT Press,  2010.

\bibitem[Nielsen(2012)]{wikiacad2}
Nielsen, F.A.
\newblock Wikipedia Research and Tools: Reviews and Comments.
\newblock {\em SSRN Electronic Jourmal} {\bf 2012}, 
\newblock {\url{http://dx.doi.org/10.2139/ssrn.2129874}}.

\bibitem[Ball(2023)]{wikiacad3}
Ball, C.
\newblock Defying easy categorization: Wikipedia as primary, secondary and tertiary resource.
\newblock {\em Insights} {\bf 2023}, {\em  7},~1.
\newblock {\url{DOI: https://doi.org/10.1629/uksg.604}}.

\bibitem[Arroyo-Machado et al.(2024)]{wikiacad4}
Arroyo-Machado, W., Diaz-Faes, A.A., Herrera-Viedma, E, Castas, R.
\newblock From academic to media capital: To what extent does the
scientific reputation of universities translate into
Wikipedia attention?
\newblock {\em J. Assoc. Inf. Sci. Technol} {\bf 2024}, {\em  75},~423.
\newblock {\url{https://asistdl.onlinelibrary.wiley.com/doi/full/10.1002/asi.24856}}.

\bibitem[Ermann e~al.(2025)Ermann, Frahm and Shepelyansky(2025)]{wiki25}
Ermann, L.; Frahm, K.M.; Shepelyansky D.L.
\newblock Competition of entries from Ising Wikipedia networks
\newblock {\em Wiki Workshop} May 22 (2025).
\newblock {\url{https://wikiworkshop.org/2025/paper/wikiworkshop_2025_paper_4.pdf}}.

\bibitem[Brin and Page(1998)]{brin}
Brin, S.; Page, L.
\newblock {The anatomy of a large-scale hypertextual Web search engine}.
\newblock {\em {Computer Networks and ISDN Systems}} {\bf 1998}, {\em 30},~107.

\bibitem[Langville and Meyer(2006)]{meyer}
Langville, A.; Meyer, C.
\newblock {\em {Google's PageRank and beyond: the science of search engine
  rankings}}; Princeton University Press: Princeton,  2006.


%
%

\end{thebibliography}
\end{document}